\documentclass[9pt,a4paper]{article}

\usepackage{setspace}

\usepackage{mathtools}
\usepackage{graphicx}
\usepackage{authblk}
\graphicspath{{../figs/}}
\usepackage[natbib]{biblatex}
\title{Reproducing scientists' mobility:\\A data-driven model}

\author[a]{Giacomo Vaccario}
\author[a,b]{Luca Verginer}
\author[a,*]{Frank Schweitzer}

\affil[a]{ETH Z\"urich, Chair of Systems Design, CH-8092 Z\"urich, Switzerland}
\affil[b]{IMT School for Advanced Studies Lucca, AXES Lab, IT-55100 Lucca, Italy}
\affil[*]{Correspondence to fschweitzer@ethz.ch}

\begin{document}

\maketitle

\begin{abstract}
High skill labour is an important factor underpinning the competitive advantage of modern economies.
Therefore, attracting and retaining scientists has become a major concern for migration policy.
In this work, we study the migration of scientists on a global scale,
by combining two large data sets covering the publications of 3.5 Mio scientists over 60 years.
We analyse their geographical distances moved for a new affiliation and their age when moving, this way reconstructing their geographical ``career  paths''.
These paths are used to derive the world network of scientists mobility between cities and to analyse its topological properties.
We further develop and calibrate an agent-based model, such that it reproduces the empirical findings both at the level of scientists and of the global network.
Our model takes into account that the academic hiring process is largely demand-driven and demonstrates that
the probability of scientists to relocate decreases both with age and with distance.
Our results allow interpreting the model assumptions as micro-based decision rules that can explain the observed mobility patterns of scientists.
\end{abstract}

\paragraph{keywords:}high-skilled labor $|$ mobility network $|$ data-driven model
\vspace{1cm}

\section*{Introduction}

Scientists are highly mobile individuals.
This has been true in the past and is even more true today~\cite{geuna2015global}.
Therefore, an increasing number of works analyses the mobility of scientists and their motivation to relocate \cite{dahl2010migration, miguelez2014what}.
Many publications have focused on the relationship between movements and scientific impact~\cite{Franzoni2014, Fernandez2015, franzoni2015international, Scellato2017, petersen2018multiscale}.
Other works have analysed scientists mobility within and across \emph{countries}, to determine policy impacts~\cite{moretti2014state, czaika2017gravity} or to study the brain circulation phenomenon~\cite{Benassy2013, Saxenian2005, Agrawal2011,verginer2020global}.

While most of these studies focus on the aggregated level, e.g., on bilateral flows between countries, there is a need to better understand scientific mobility at the individual level~\cite{appelt2015which, fortunato2018}.
Empirical works in this direction ~\cite{gibson2014scientific, veugelers2015destinations} are often based on survey data that provide only partial coverage of the global mobility of scientists.
Theoretical works on scientist mobility~\cite{Mahroum2000}, on the other hand, are rarely validated against data.
Researches in complexity and network theory have mostly analyzed scientific collaborations \cite{newman2001structure,wagner2005network,gomez2019clustering} or hiring practices \cite{clauset2015systematic}.
%, and modeled innovation processes~\cite{gilbert2001innovation, konig2011recombinant, monechi2017waves}.
%Still, models and empirical findings on scientist mobility at individual and especially city level are scant.

Our work addresses this research gap in a two-fold manner.
First, we provide empirical insights into scientific mobility at the individual level, by reconstructing 3.5 million geographical career paths from large-scale data sets.
Second, we provide an agent-based model that is calibrated against the available data and is capable of reproducing the distributions of relocation \emph{distances} and relocation \emph{age}.
% However,
In developing our \emph{data-driven} model, we follow the approach of~\cite{tomasello2014role, tomasello2017data, vaccario2018quantifying}.
Our model incorporates two factors that are known to affect academic mobility: (i) geographical distance~\cite{morgan2004exaggerated,miguelez2014what}, (ii) prestige, and selectiveness of academic institutions~\cite{clauset2015systematic,vaccario2020,Verginer2020talent}.
We also contribute to the understanding of global mobility, by reconstructing and analysing the world network of scientists mobility at the level of \emph{cities}, not countries.
From this global network, we extract topological features such as the distributions of degrees, local clustering coefficients, path lengths, and assortativity, to demonstrate that these can also be reproduced by our agent-based model.

% %Our model, together with its calibration and validation procedure, is explained in the Sect.~\ref{sec:model-mobil-scient}.
% Finally, we discuss the results from our simulations and analyse the limitation of the model. % and provide some outlooks.

\section*{Results}

\subsection*{Empirical findings}

By combining two large-scale bibliographic datasets as described in \emph{Materials and Methods}
we obtain for $N= 3' 740' 187$ scientists information about the sequence of cities they worked in their careers, between 1950 and 2009.
The merged dataset contains $M=5' 485$ unique cities.
The data allows us to construct the geographical ``career path'' of these scientists.
An illustrative example is given in Table~S1. %\ref{tab:example_medline_record} in the SI.
\begin{figure*}
       \centering
       \footnotesize
       (a)\includegraphics[width=0.28\textwidth]{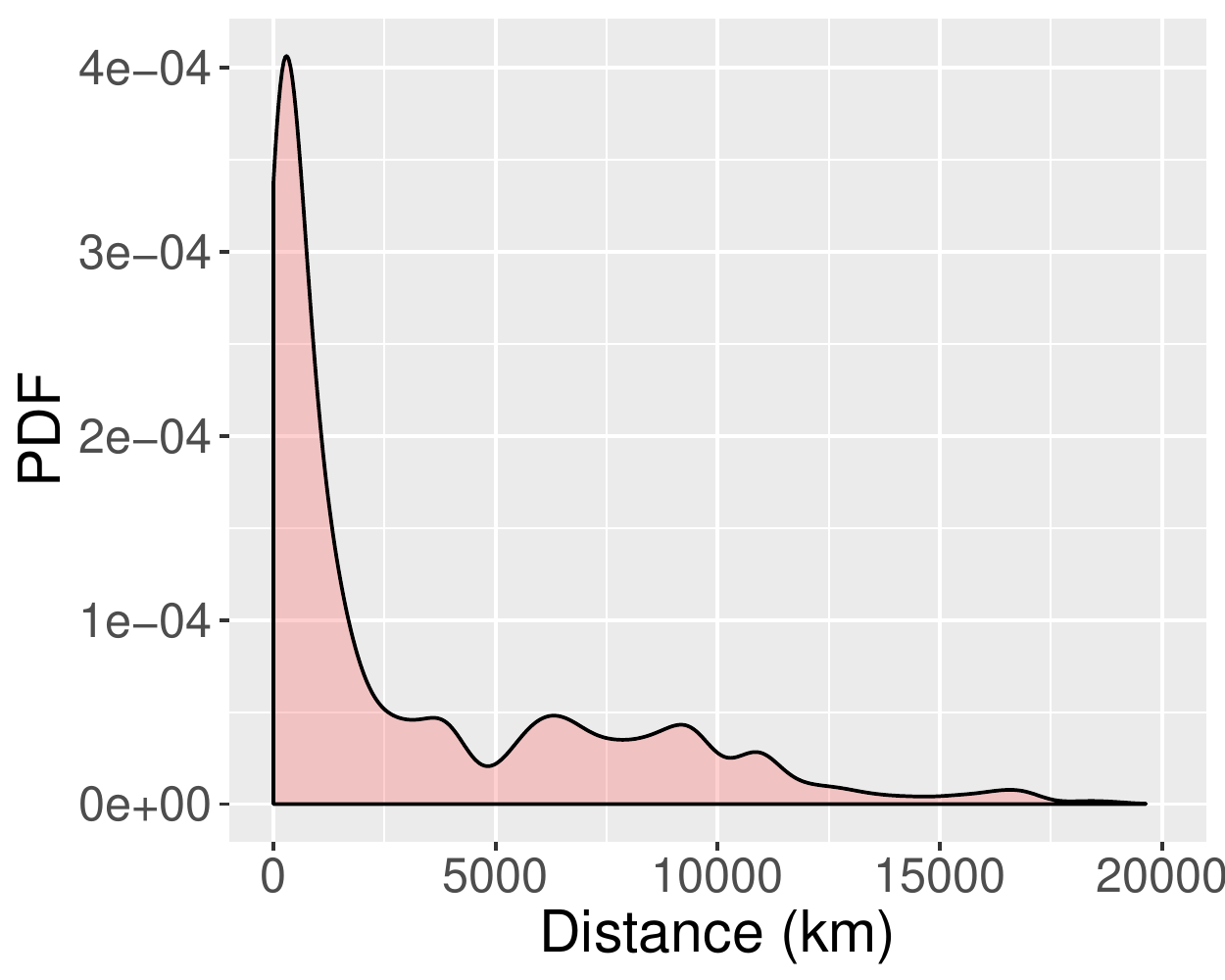}
       \hfill
       (c)\includegraphics[width=0.30\textwidth]{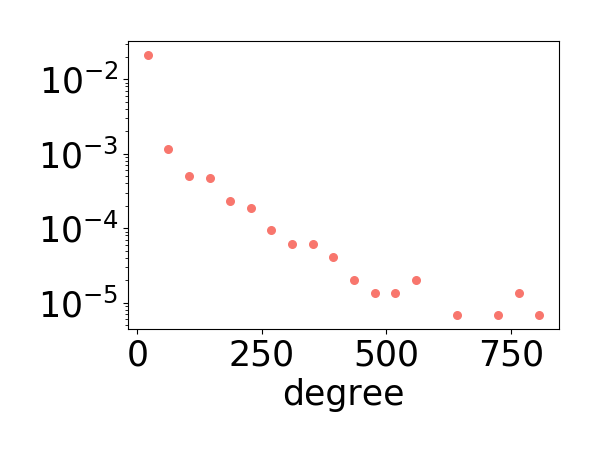}
       \hfill
       (e)\includegraphics[width=0.30\textwidth]{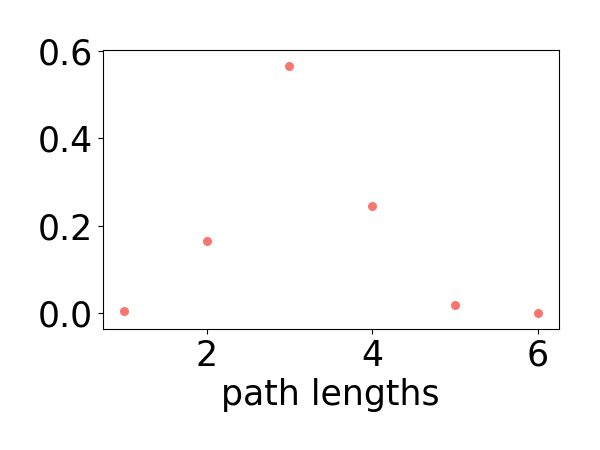}
       \\
       (b)\includegraphics[width=0.28\textwidth]{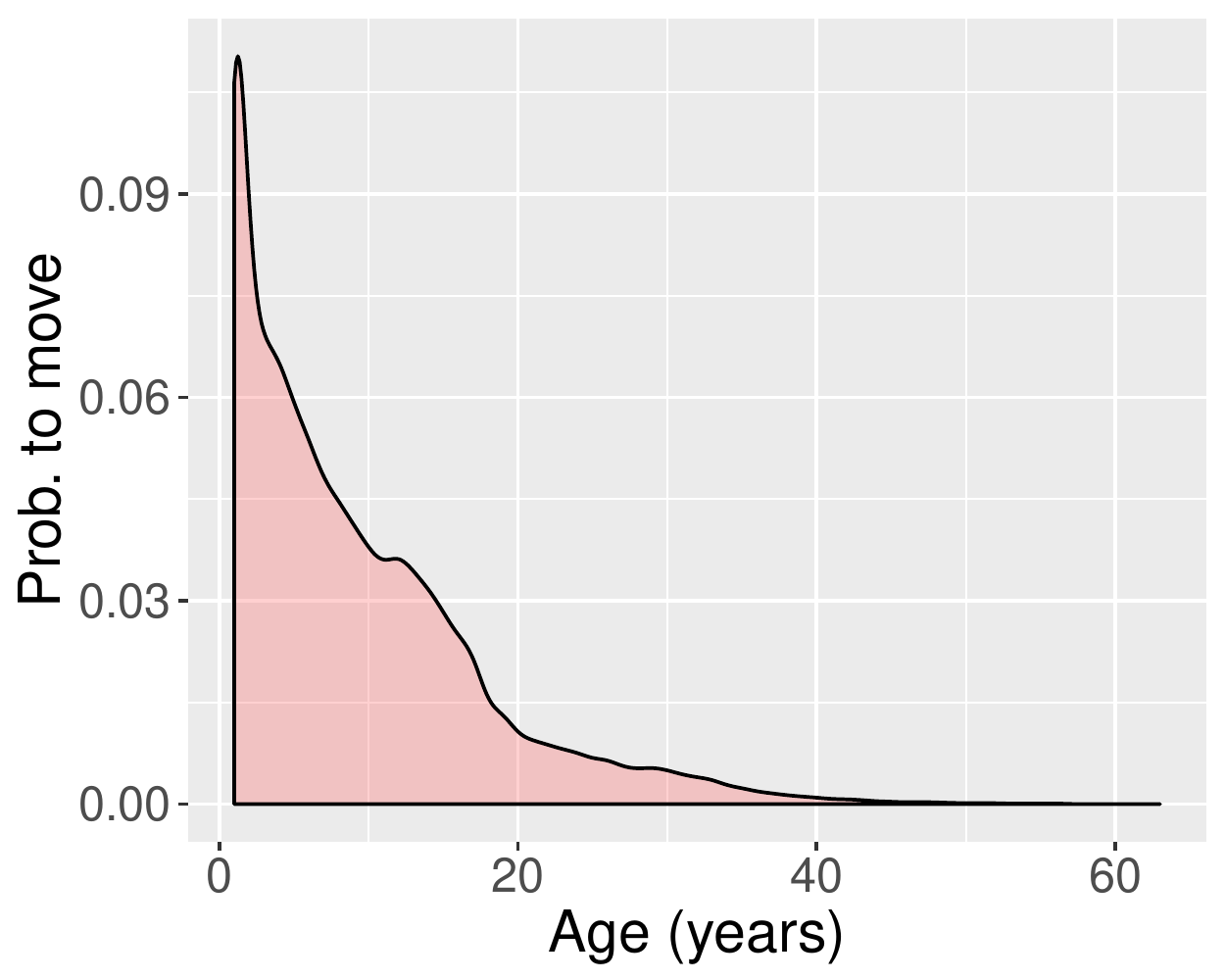}
       \hfill
       (d)\includegraphics[width=0.30\textwidth]{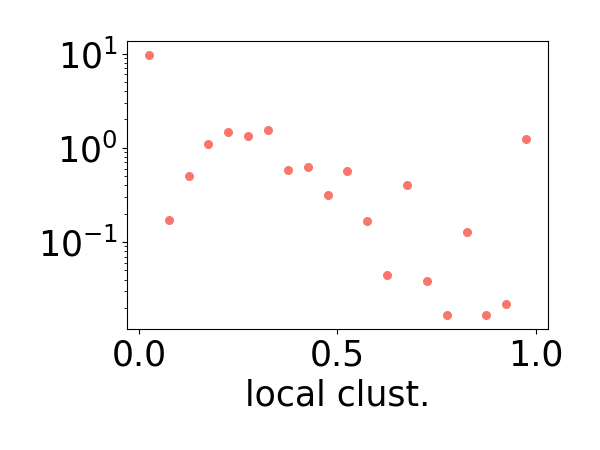}
       \hfill
       (f)\includegraphics[width=0.30\textwidth]{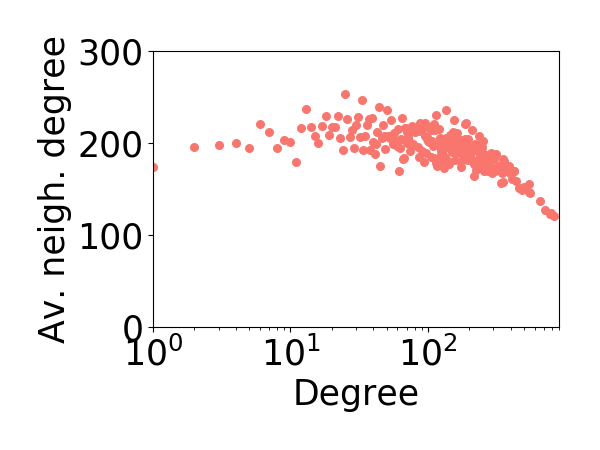} \\
       \vspace{0.5cm}
       (g)\includegraphics[clip, trim = 100 230 100 100, width=.98\textwidth]{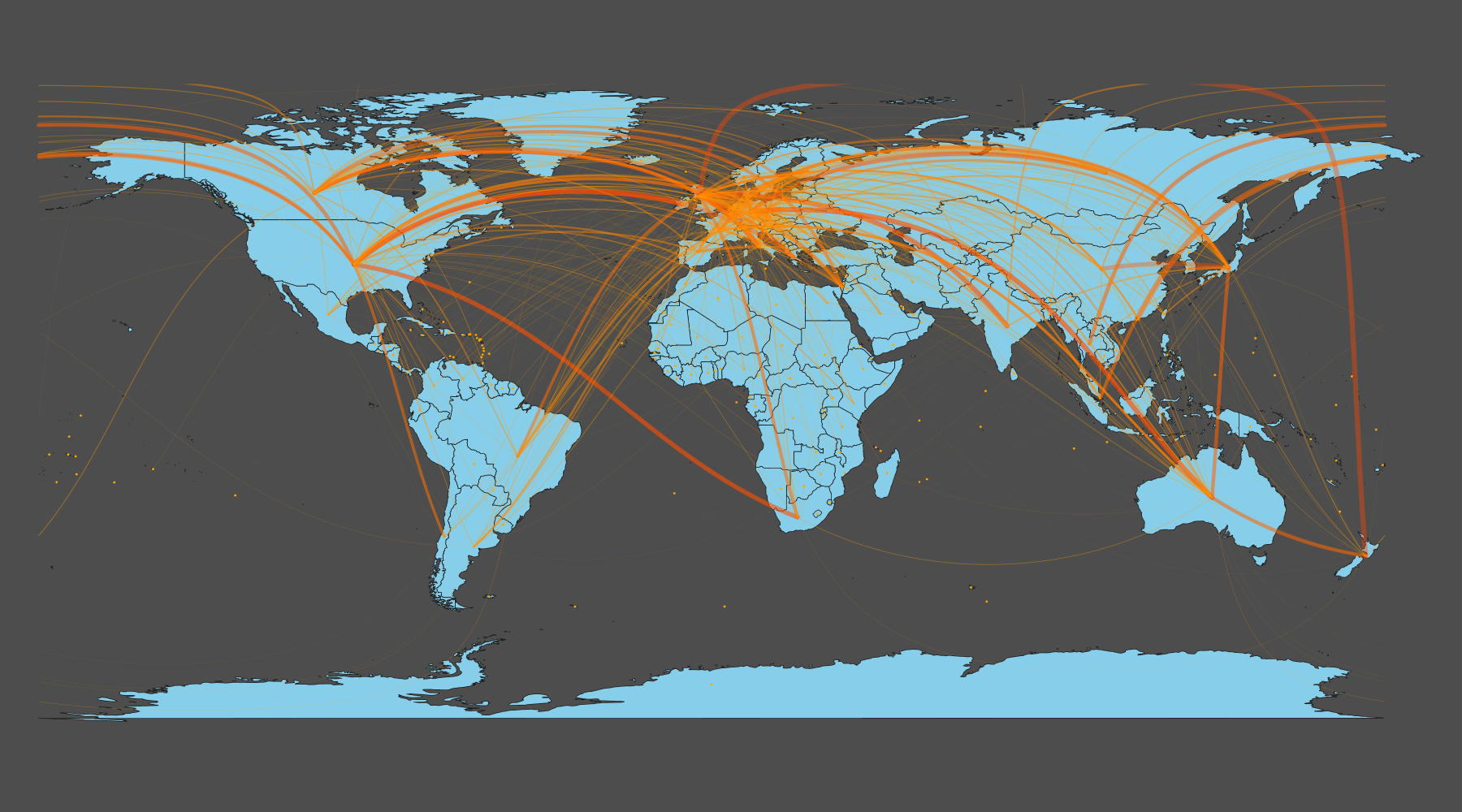}
       \caption{\textit{Characterization of the empirical academic mobility}.
       At the individual level, we have the distribution of relocation distances of scientists (a) and the distribution of moves dependent on the (academic) age of scientists (b).
       At the global level, we reconstruct the mobility network for which, we plot the distribution of degrees (c), local clustering coefficients (d) and path lengths (e).
       In (d) we plot the average degree of neighbors of a node with degree $k$ in function of $k$.
       Finally, the mobility network of scientists in between 1990 and 2008 in (g).
       The link width and the color indicate the magnitude of the \emph{total flow} between any two countries.
       For visualization purpose, the total flows have been aggregated at country level and logarithmically scaled.
       }\label{fig:emp-prop}
   \end{figure*}

\paragraph{Statistics of geographical career paths. }

From the career paths we compute the \emph{relocation distances} scientists moved when changing their affiliation, using the Haversine formula for geodesic distances.
   % if we consider only France Germany and the UK
   % The distribution obtained from 7663 authors is shown in Figure~\ref{fig:traj-valid}(a).
   % We note that it is a left-skew distribution with a median of 278 km.
   % I.e. most scientists find a new affiliation in cities within a radius of 300 km around their current affiliation.
   % However, movements of more than 500 km toward distant cities are also quite frequent.
   %if we consider the full world (see images figs/pdf_of_paths_distance_inkm_empirical_entire_world.pdf
   % and figs/pdf_tomove_given_age_empirical_entire_world.pdf for the age distribution)
The distribution obtained from $62\,465$ scientists relocating between 2000 and 2008 is shown in Figure~\ref{fig:emp-prop}(a).
We note that it is a left-skew distribution with a median of $1\,000$ km, i.e., most scientists find a new affiliation within a radius of $1\,000$ km around their current affiliation.
However, relocations toward cities that are more than $6\,000$ km away are also quite frequent.

The data also allows us to relate the frequency of such moves to the age of scientists.
Because the physical age of scientists is unavailable, we rely on their \emph{academic age}, $t^{a}_{i}$, also measured in years.
$t_{i}^{a}=0$ when the scientist publishes his/her first paper, according to our database. % (which is probably a physical age of about 25 years).
The frequency of any recorded moves over the academic age $t^{a}$ is shown in Figure~\ref{fig:emp-prop}(b).
Again, it is a left-skew distribution with a median of 7 years.
This matches the known fact that the mobility of scientists drastically decreases with age~\cite{canibano2011international, verginer2018brain}.
However, we also identify that some scientists change their working location
% , regardless of having an academic age higher than 40 years, i.e.,
after been active for 40 years. %  (which is probably at the physical age of retirement).

% We define a ``move'' on a path $p_i$ given a ``move date'' $t_M$ and a buffer ($b$) around it as follows.
% Take all locations passed through in the first window $[t_M-b, t_M)$ and sort the locations by observations time (i.e. $A,A,B$).
% Within the window chose as the source location the location with the longest uninterrupted sequence, in case of a tie chose the one closest to $t_M$ (i.e. $A$).
% Repeat the same procedure for the second period $[t_M+b)$, which in our example yields the move $(A,C)$.
% Repeating this procedure for a given $t_M$ and $b$ for all author paths yields a Mobility Network ($M_{t_M}$) showing how many authors moved from a source location to a target location within the given time interval.

\paragraph{Reconstructing the mobility network of scientists. }

While the career paths and their statistics refer to individual scientists, we can also analyse the network that results from aggregating all of the career paths of a given year.
This aggregation changes the unit of analysis to the \emph{city level}.
For each year, we obtain the number of scientists $N_{K}(t)$ in a given city $K$ from their publications by taking unique geo-located scientists into account.

We further calculated for each year $t$ the number of scientists $\Delta N_{K\gets L}(t)$ moving into city $K$ from another city $L$, i.e., the inflow, and the number of scientists $\Delta N_{L\gets K}(t)$ moving out of city $K$ to another city $L$, i.e. the outflow.
%    \begin{figure}
%        \centering
%        \includegraphics[width=0.24\textwidth]{pdf_activity_source_cities_log_empirical_entire_world}
%        \includegraphics[width=0.24\textwidth]{pdf_activity_target_cities_log_empirical_entire_world}\\
%        \caption{Distributions of (a) inflow of scientists into any city, (b) outflow of scientists out of any city.
%        The $x$-axis is in log-scale.
%        }\label{fig:in-out-flow}
%    \end{figure}
%    Figure~\ref{fig:in-out-flow}(a) and (b) show the respective distributions for the aggregated inflow $\Delta N_{K}^{\mathrm{in}}(t)=\sum_{L}\Delta N_{K\gets L}(t)$ of scientists into city $K$ and the aggregated outflow $\Delta N_{K}^{\mathrm{out}}(t)=\sum_{L}\Delta N_{L\gets K}$ of scientists out of city $K$.
%    The aggregate in- and the outflow are computed at three different time windows centred at 2000, 2002 and 2004.
%    This means that each city is considered three times (once in each time window).
%    Again, we note the left-skew distribution for both quantities, which indicates the \emph{heterogeneous} contribution of cities to the global movement of scientists.
For any given pair $(K, L)$ of cities, we then calculate the \emph{total flow} of scientists between these two cities,
%The total flow is the total number of scientists exchanged between $K$ and $L$,
$\Delta N_{L\gets K} + \Delta N_{K\gets L}$.
This flow allows us to visualise the mobility network of scientists at the world level, shown in Figure~\ref{fig:emp-prop}(g).
The links are undirected but weighted according to the total flow.

%    \begin{figure}
%        \centering
%        \includegraphics[clip, trim = 100 230 100 100, width=.48\textwidth]{world_migration.png}
%        \caption{The mobility network of scientists in between 1990 and 2008.
%        The link width and the colour indicate the magnitude of the \emph{total flow} between any two cities.
%        For visualisation purpose, the total flows have been aggregated at country level and logarithmically scaled.}\label{fig:world_migration}
%    \end{figure}

\paragraph{Topological properties of the mobility network. }

To obtain the topological properties common in network analysis, we aggregate the mobility networks for the period 2000 -- 2008.
On this \emph{aggregated network}, we calculate standard measures, such as the \emph{degree distribution} $P(d)$, where $d$ is the number of cities scientists in a given city either move to or come from.
Figure~\ref{fig:emp-prop}(c) shows that this is a broad distribution.
Some cities act as hubs, with a large degree, most cities, however, only have a small degree.%, which is also visualised in  Figure~\ref{fig:emp-prop}(g).

The distribution of \emph{path lengths}, shown in Figure~\ref{fig:emp-prop}(e), measures how many steps are needed to reach, on the network, any city from a given starting point.
The small number of steps indicates that the network is dense in a \emph{topological sense}, not necessarily in a geographical one.

The \emph{local clustering coefficient}, on the other hand, measures whether three neighbouring cities (with respect to their geographical proximity) form closed triangles, i.e., whether there is an exchange of scientists among all three of them.
Figure~\ref{fig:emp-prop}(d) shows the distributions of these values, and we find that most cities have a small local clustering coefficient.
%which is in the world map of Figure~\ref{fig:world_migration} shown as a densely connected area.

The \emph{neighbor connectivity}, eventually, measures to what extent cities with a certain degree are connected to other cities with a similar degree.
Figure~\ref{fig:emp-prop}(f) shows a non-monotonous dependency.
Cities with a low degree tend to show an \emph{assortative} pattern, i.e., they are connected to cities that have a similar number of neighbours.
Cities with a high degree, which are characterised as \emph{hubs} above, are rather connected to cities with a lower degree, i.e., they are \emph{dissortative}.
This result gives us already on the topological level important information about the origin of scientists coming to the hubs and the destination of scientists leaving the hubs.
They do not hop between hubs, which would have been indicated by an assortative pattern.

% This is confirmed by the degree distribution shown in Figure~\ref{fig:net-valid_1}(c).
% To investigate how clustered are the cities, we report in Figure~\ref{fig:net-valid_1}(a) the distribution \emph{local clustering coefficient}.
% The local clustering coefficient of a node measures how connected neighbours are its neighbouring nodes.
% We find that the distribution is picked at quite small values.
% The presence of hubs and the small clustering coefficient of nodes suggest that the network might be easily traversed.
% We validate this intuition by studying the \emph{path lengths},
% i.e. how many stops are needed to reach, on the network, any city from a given starting point.
% In Figure~\ref{fig:net-valid_1}(b), we plot the distribution of path lengths.
% The small number of hops confirms the intuition that the network can be traversed very easily.
% This means that any city can be easily reached from a given starting point.

%    \begin{figure}
%        \centering
%        \includegraphics[width=0.24\textwidth]{transitivity}
%        \hfill
%        \includegraphics[width=0.24\textwidth]{paths}\\
%        \includegraphics[width=0.24\textwidth]{degree}
%        \hfill
%        \includegraphics[width=0.24\textwidth]{degree_vs_knn} \\
%        \caption{Distributions of (a) local clustering coefficients, (b) path lengths and (c) degrees.
%        In (d) we plot the average degree of neighbours of a node with degree $k$ in the function of $k$.
%        }\label{fig:net-prop}
%    \end{figure}

\subsection*{Modeling the mobility of scientists}

We now develop a model to reproduce the characteristic empirical properties of the scientists' mobility network discussed above.
Precisely, we want to reproduce features both at \emph{scientist} and \emph{network} level.
These are, at the scientist level, (i) the distribution of moved distances, Figure~\ref{fig:emp-prop}(a) and (ii) the ``age at move'' distributions, Figure~\ref{fig:emp-prop}(b).
At the network level, we want to reproduce (iii) the distributions of the topological features shown in Figure~\ref{fig:emp-prop}(c-f), i.e., degrees, local clustering coefficients, path lengths, and average neighbour degree.

We develop an \emph{agent-based model} because we want to model the relocation of \emph{scientists}, as opposed to a system dynamics model that would merely reproduce the flows between \emph{cities}.
This choice implies that \emph{macroscopic features} describing the system, such as the topological properties already discussed, must be \emph{emergent properties} arising from the agent dynamics.

Our model is composed of two entities, \texttt{agents} and \texttt{locations}.
\texttt{Agents} represent scientists.
Each agent $i$ is characterized by three properties that change over time: its position, ${r}_i(t)$, its fitness, $f_i(t)$, and its years of activity $y_i(t)$.
Time is measured in discrete simulation steps, each step representing one year.
When we start our simulations at time $t=0$, which is chosen as the year 2000 below, we have to consider that many
agents have already published before 2000, which is included in $y_{i}(t)$.
For instance, an agent that published its first paper in 1995 will have a $y_{i}(2000)=5$.
This information is essential to measure an agent's fitness, $f_{i}(t=0)$, which we do below.

\texttt{Locations} represent cities and host agents.
%In agreement with the dataset, we have $M=5,485$ different locations.
Each location $K$ is characterised by three properties that can also partly change over time: its position ${R}_{K}$ defined in real geographical space by means of longitude and latitude, its fitness, $F_{K}(t)$, and the number of agents it hosts, $N_{K}(t)$.
Note that ${R}_{K}$ and $N_{k}(t)$ are taken from the available empirical data.
For the position ${r}_i(t)$ of an agent, we assume that at each time step the agent can be found in one of the available locations.
So ${r}_i(t)=R_{K}$ where $K$ is the location, where agent $i$ is located at time $t$.

\paragraph{Agent and location fitness.}
The individual agent fitness $f_i(t)$ represents the academic impact of a scientist.
We proxy this impact by the papers that he/she has co-authored.
Precisely, we assign to each paper a score equal to the impact factor of the journal (taken from \textsc{SCImago}) where it was published divided by the number of co-authors.
Then, for each scientist, we aggregate the scores of his/her co-authored papers in the last two years of activity.
By this, we obtain a distribution of fitness values that we can assign to agents.

We assign to each location $K$ a fitness value $F_{K}(t)$ reflecting the quality of the academic institutions hosted in a city.
To make this idea explicit and measurable, we take $F_{K}(t)$ to be the mean fitness of the agents located in $K$.
Note that this approach is in line with how rankings of academic institutions are created.
Indeed, university rankings are determined considering the academic impact and quality of the scientists working there.
In our model, we assume that the $F_{K}(t)$ is public information, and thus, agents may use this information in their decision rule.

\paragraph{Relocation preferences.}
Our central modelling assumption is that agents prefer to work in locations that provide a higher fitness than the one they are currently based.
These locations, however, can be distant from the current location, which implies higher relocation costs.
Therefore, an agent $i$ takes into account the fitness $F_{K}(t)$ of locations and its geodesic distance $\Delta_{i, K}(t)$.
Agents combine this information in a \emph{re-scaled fitness score} $\tilde{F}_{i,K}(t) = F_{K}(t) / (\Delta_{i, K}(t))^b$ for each location $K$.
$b$ is a model parameter, used to weight the impact of spatial distances.
The bigger $b$, the more important any spatial distance becomes.
%% NOTE the term friction is wrong. It comes from physics and has a different meaning. -- is any bigger (smaller) is the spatial friction.

Ranking the values $\tilde{F}_{i, K}(t)$ from high to low, each agent then obtains an \emph{individual} ranking that reflects its preferences where to move to.
Agents in $L$ will consider only those locations where ${F}_{K}(t)>F_{L}(t)$, i.e., where the average fitness of scientists is larger than the average fitness of scientists in their city.
Hence, each agent $i$ assigns to a location $K$ the score:
\begin{equation}\nonumber
    R(i,K) =  \Theta\left[ {F}_{K}(t) - F_{L}(t)\right] \frac{F_{K}(t)}{(\Delta_{i, K}(t))^b}
\end{equation}
where $\Theta\left[ {F}_{K}(t) - F_{L}(t)\right]$ is equal to 1 when ${F}_{K}(t)>F_{L}(t)$ and equal to 0, otherwise.
% This captures our assumption that scientists try to move to cities hosting scientists with average higher academic impact.
% For each agent-$i$ we create a ranking vector of all locations depending on a
% where $\Delta_{i,K}(t)$ is a geodesic distance between location-$K$ and agent-$i$ at time $t$, and

%The fitness $f_i(t)$ is an intrinsic quality of agent-$i$ at time $t$.
% This captures the academic impact of a scientist.
% The years of activity $y_i(t)$ is the number of simulation time steps during which agent-$i$ was active.
% This represents the academic age of a scientist, i.e. the number of years elapsed from her/his first publication.
% Agents are hosted in locations.

%The position ${R}_{K}$ is a vector with the geographical coordinates of location-${K}$.

% Under the assumption that scientists prefer to work in cities where the average academic impact is high,
% the fitness $F_{K}(t)$ represents the attractiveness of a city from a scientist's perspective.
% %By this we assume that.
% The size $N_{K}$ is the maximum number of agents that location-$K$ can host.
% This represents the observed number of scientists in a city.

% During each simulation time step,% agents move between locations.
% %Precisely
%  we try to relocate agents to new locations with higher fitness.

\paragraph{Relocation decisions.}
Agents only come up with a ranked list of possible locations they would consider to move to (and we assume that they send applications to the academic institutions in these locations).
However, agents do not decide where to move.
This decision, whether or nor to accept the agent, is taken by the location.

\begin{figure}
    \centering
    \includegraphics[width=.8\textwidth]{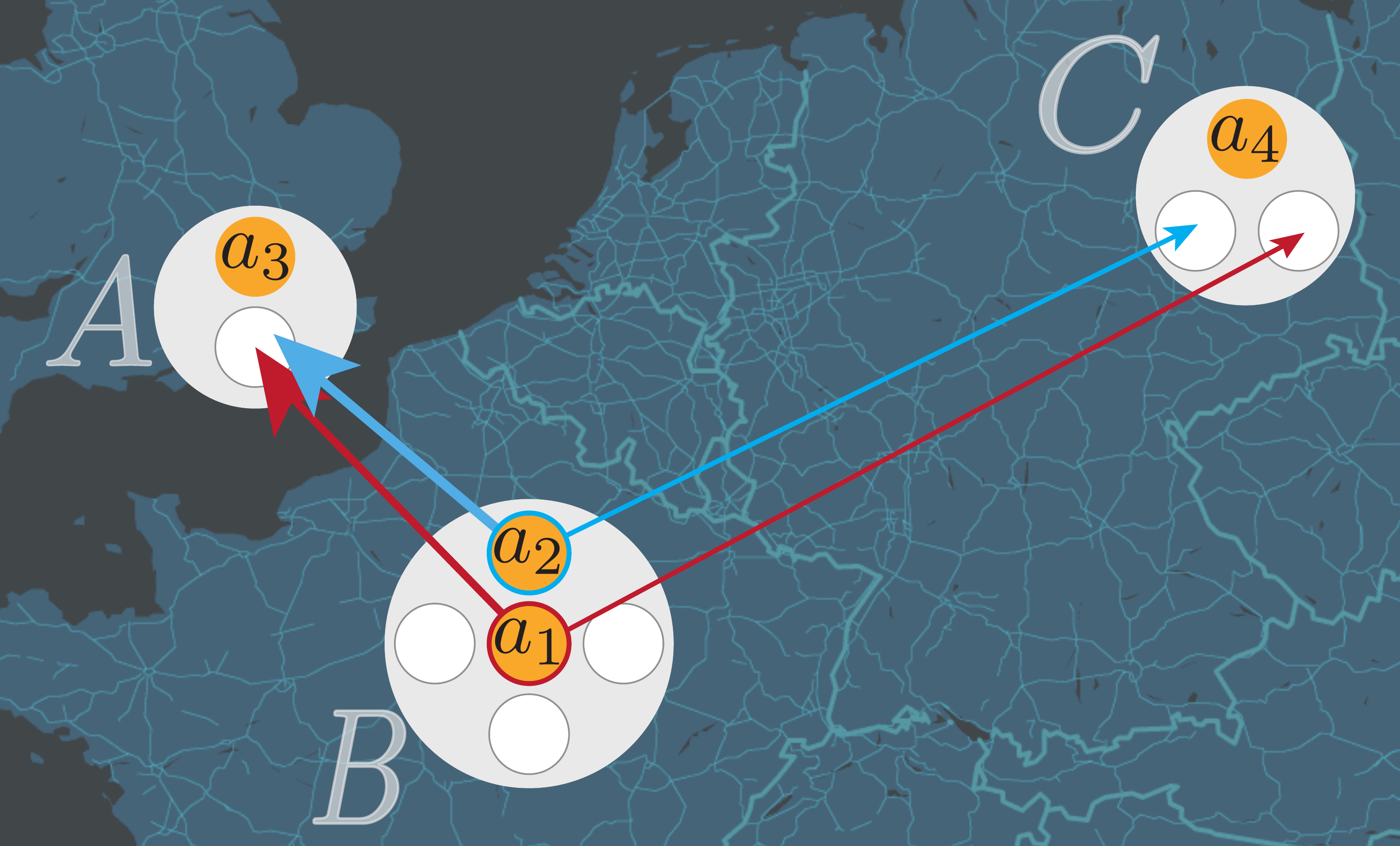}
    \caption{\textit{Example of relocation rules}.
    Four agents ($a_1$, $a_2$, $a_3$ and $a_4$) are hosted in three locations, $A$, $B$ or $C$, representing London, Paris and Berlin.
    Each location has a maximum number of available positions illustrated by small circles: $N_A = 2$,  $N_B = 4$ and $N_C = 3$.
    In this image, agents $a_1$ and $a_2$ compute the rescaled fitness of the available locations ($A$ and $C$) and rank these location accordingly.
    Here, we have assumed that $A$ and $C$ have the same fitness ($F_A(t) = F_C(t)$),
    but $A$ is closer to $B$ than $C$ ($\Delta_{i,A} < \Delta_{i,C}$ for $i=1,2$).
    For this reason, both $a_1$ and $a_2$ express a preference for $A$ over $C$.
    Since location $A$ has $N_A = 2$ and one position is already taken,
    $A$ must decide to accept either $a_1$ or $a_2$, depending on their fitness.
    }\label{fig:abm_illustration}
\end{figure}

A location $K$ will accept new agents only if it has sufficient capacity.
The capacity $N_{K}(t)$ for a given city $K$, is estimated from the number of scientists empirically observed in city $K$ in year $t$.
% External factors, such as the growth of academic institutions, are implicitly considered in the observed change of $N_{K}(t)$.
%The $N_{K}(t)$ are stable over the period 2000-2006.
%For this reason, we do not model city growth.
% This implies that, after some transient periods in our simulations, locations have the capacity to accept incoming agents only if agents at $K$ have been accepted somewhere else and move there.
   Dependent on the individual ranking of agents, some locations obtain more applications than the capacity allows them to accept.
So each location ranks the qualified agents according to their fitness $f_{i}(t)$.
Available slots are filled starting from agents with higher fitness values until the capacity $N_{K}(t)$ is reached.
Precisely, if $f_i(t)>F_K(t)$, location $K$ considers agent $i$ with probability $p=1$ because this allows location $K$ to increase its fitness $F_{K}(t)$.
If $f_i(t)\leq F_K(t)$, location $K$ considers agent $i$ only with a probability $p = (f_i(t)/F_{K})^s$ where $s$ is our second model parameter.
This parameter $s$ represents the \emph{selectiveness} of locations, the higher $s$, the more difficult it is to be hired.
Hence, if a location $K$ has some openings, its probability to accept agent $i$ is:
\begin{equation}\nonumber
    p(K,i) = \begin{cases}1 &  f_i(t) > F_{K}(t)\\
    \left(\frac{f_i(t)}{F_{K}(t)}\right)^{s} & \text{otherwise} \end{cases}
\end{equation}
%    \begin{figure}
%        \footnotesize
%        (a)\includegraphics[width = 0.30\textwidth, clip, trim =25 10 24 20]{pki1.pdf}
%        (b)\includegraphics[width = 0.30\textwidth, clip, trim =25 10 24 20]{pki2.pdf}
%        (c)\includegraphics[width = 0.30\textwidth, clip, trim =25 10 24 20]{pki4.pdf}
%        \caption{Probability of a location $K$ to accept an agent $i$ in function of the ratio between the agent and location fitness.
%        In we set $s=0.5$, 1, 5 respectively in panel (a), (b), and (c).}
%        \label{fig:decision-prob}
%    \end{figure}
%    In Figure~\ref{fig:decision-prob}, we plot the acceptance probability $p(K,i)$ for three different values of the parameter $s$.
%    Note that for big value of $s$, locations becomes more selective (see Figure~\ref{fig:decision-prob}(c)) as almost only agent with $f_i(t)\geq F_{K}$ are accepted with high probability.
%    For smaller values of $s$, locations becomes less selective (see Figure~\ref{fig:decision-prob}(a)).
In Figure~\ref{fig:abm_illustration}, we summarise and visualise the basic rules of our model.

\paragraph{Matching agents to locations. }

In our model agents rank locations, while locations rank agents.
To match locations and agents, we have to solve a matching problem similar to the stable marriage problem.
However, our problem is slightly different as a location can accept more than one agent until the capacity $N_{K}(t)$ is reached.
%Each location-$K$ can host new agents until the number of hosted agents reaches $N_K$.
To solve this matching problem, we use an applicant proposing algorithm, similar to the NRMP-algorithm~\cite{roth1999redesign}.
The details are given in the \emph{Materials and Methods} section.

\paragraph{Fitness dynamics. }

To model those agents not accepted at a new location, we consider that agent which stay at their current location, i.e., $r_i(t+1) = r_i(t)$, use the time step to further improve their fitness, $f_{i}(t)$.
For this we assume a stochastic dynamics, precisely an additive stochastic process with a variance proportional to the fitness of the current location:
\begin{equation}\nonumber
    f_{i}(t + 1) = f_{i}(t) + \alpha_{\mathrm{loc}} \eta
\end{equation}
where $\alpha_{\mathrm{loc}}$ is a parameter proportional to the quality of the agent location, and $\eta$ is a normally distributed stochastic variable with 0 mean and variance equal to 1 (i.e.,$\eta\approx\mathcal{N}(0,1)$).
By this, we assume that the change in fitness of an agent depends on its location.
Also, it is not guaranteed that agents will improve their fitness; they can also lower it.

At the end of each time step, we update the fitness of locations, $F_{K}(t)$, by averaging over the fitness $f_{i}(t)$ of all those agents that are currently based there.

\paragraph{Entry and exit of agents. }
At every time step, agents can exit and enter the system.
This dynamic simulates the fact that academia is an open system, i.e., every year, scientists exit the systems, but also new ones enter it.
Precisely, agents are removed with a small probability of $p_e$ at the end of every time step.
Hence, the probability that an agents is removed from the simulation after $n$ steps is $(1-p_e)^{n-1}p_e$.
This process allows us to replicate the observed (academic) survival probability function of scientists (see Figure~S2 in SM).

\begin{figure}
  \centering
  % \includegraphics[width=0.18\textwidth, clip, trim = 0 0 86  0]{pdf_of_paths_distance_inkm_validation}
  % \hfil(b)
  % \includegraphics[width=0.25\textwidth, clip, trim = 0 0 0 0]{pdf_tomove_given_age_by_year_validation}
  \includegraphics[width=0.8\textwidth]{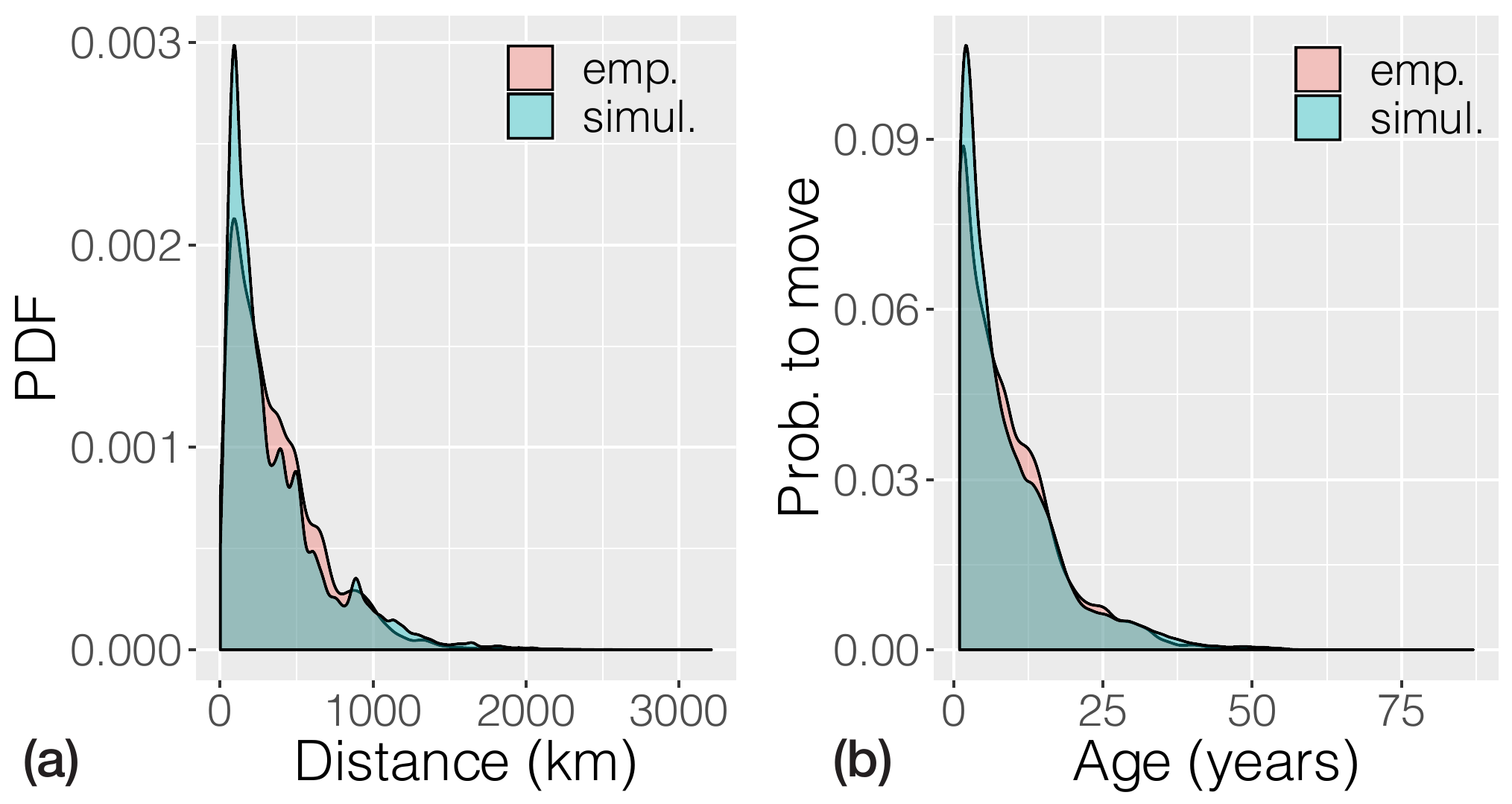}
  \caption{
  \textit{Comparison of simulated and empirical individual scientist features.}
  (a) Distribution of relocation distances of scientists.
(b) Distribution of moves dependent on the (academic) age of scientists.
(red) indicates the empirical distribution, (blue) the distributions that are obtained from our agent-based simulations.
% The distributions are obtained from the frequencies using a kernel density estimation.
%      travelled by scientists when relocating for the empirical and the simulated network.
%      Similarly in (b) we show the correspondence of age distribution between the empirical and simulated mobility networks.
}
\label{fig:traj-valid}
\end{figure}

\paragraph{Model calibration.  }

We use the empirical data not only as an \textit{input} to our model, to determine the \emph{initial conditions},
but also for \textit{calibration}.
For this, we use only a subset of the available data (see the \emph{Materials and Methods} section).
A major effort was spent to determine the optimal values of the two free parameters of our model, $b$ and $s$.
%with the calibration procedure described in the Material and Methods section and SI Appendix.
$b^{\mathrm{opt}}=0.5$ means that a location $A$ which is twice as far away as $B$ must have a fitness $F_A=F_B * \sqrt{2}$ to be as attractive as $B$.
Moreover, $s^{\mathrm{opt}}=0.5$ means that agents are accepted by locations with a probability larger than their fitness ratio.
For example, an agent with fitness ratio $f_i/F_k=0.5$ is accepted with probability 0.71 and
an agent with $f_i/F_k=0.25$ is accepted with $p = 0.5$.

\paragraph{Model validation.  }
% and \textit{validating} it.
% We take six observed quantities into account, three at the city level and three at the scientist level,
 The calibrated agent-based model has to prove its evidence in that it can reproduce the whole set of empirical findings that have \emph{not} been used during the calibration procedure.
We \emph{validate} the model by comparing two distributions on the level of scientists, and four distributions on the level of the mobility network.
To simulate a large number of realisations, we focus on three neighbouring countries in Europe, namely Germany, France and the UK.
Furthermore, we restrict the simulation to the period 2000 to 2006.
The upper limit 2006 is given by the fact that the last publication in \texttt{Author-ity} is in 2009, and we require a 3-year window to identify moves.

\paragraph{Results of agent-based simulations. }
\label{sec:results}

\begin{figure}
    \centering
    % \includegraphics[width=0.24\textwidth]{transitivity_comparison}
    % \hfill
    % \includegraphics[width=0.24\textwidth]{paths_comparison}\\
    % \includegraphics[width=0.24\textwidth]{degree_comparison}
    % \hfill
    % \includegraphics[width=0.24\textwidth]{knn_comparison} \\
  \includegraphics[width=0.8\textwidth]{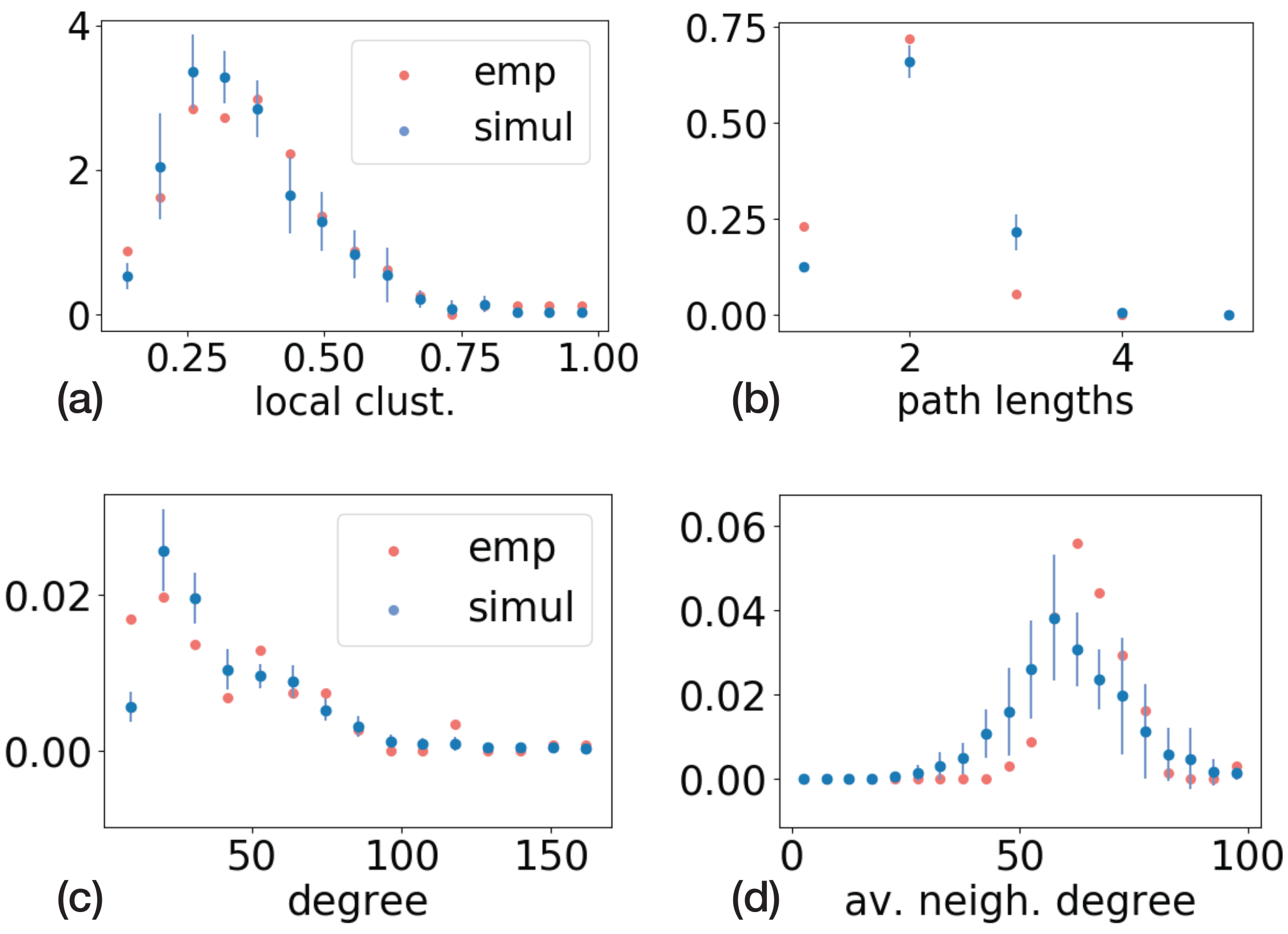}
    \caption{\textit{Comparison of empirical and simulated topological properties of the mobility network}. Distributions of (a) local clustering coefficients, (b) path lengths, (c) degrees, and (d) degrees of neighbors.
      (red) indicates the empirical distribution, (blue) the distributions that are obtained from our agent-based simulations.
      The error bars correspond to the standard deviations of the measures computed on the 10 different realisations of the simulated mobility network.
    }\label{fig:net-valid_1}
\end{figure}

The results of the validation are shown in Figure~\ref{fig:traj-valid} and~\ref{fig:net-valid_1}.
To allow for a direct comparison, we plot the empirical data in red and the simulation in blue.
We can report a good match of all distributions both on the level of scientists and on the network level.
Specifically, on the scientists' level, we are able to reproduce the two distributions of \emph{relocation distances} and of \emph{age} when moving, see Figure~\ref{fig:traj-valid}(a,b).

% the model, we verify that it reproduces four empirical distributions, such as the distribution of distances travelled by scientists.
% %We considers two scientist-centric distributions: the distribution of distances travelled by scientists,
% %Also, we consider to system-centric distributions.
% %In particular, by adopting a network perspective,
% %we look at the distribution of shortest path lengths and of local clustering coefficients.
% Note that the four distributions used for validation capture both scientist-centric and system properties not used as an input or to calibrate the model.

%\subsection*{The validation procedure}
% \begin{table}[]
%     \centering
%     \begin{tabular}{|rl|}
%         \hline
%     \multicolumn{2}{|c|}{\textbf{Reproduced distributions}}      \\
%     \hline
%     Network            & \begin{tabular}[c]{@{}l@{}}PDF of clustering coefficients (Fig.~\ref{fig:net-valid_1} (a))\\
%                                                     PDF of shortest path lengths (Fig.~\ref{fig:net-valid_1} (b))
%         \end{tabular}                               \\
%     \hline
%     Scientists-centric & \begin{tabular}[c]{@{}l@{}}PDF to observe a scientists to move given its age (Fig.~\ref{fig:traj-valid} (a)\\
%                                                     PDf of distances travelled by scientists (Fig.~\ref{fig:traj-valid} (b))\end{tabular} \\
%     \hline
%     \end{tabular}
%     \caption{Summary of the distribution used for validation.}
%     \label{tab:distr-valid}
% \end{table}

On the network level, we are able to reproduce the four distributions of \emph{clustering coefficients}, \emph{path lengths}, \emph{degree} and \emph{average neighbor degree}, see Figures~\ref{fig:net-valid_1}(a-d).
We emphasise that these results are far from being trivial.
As we start with an agent-based perspective, the results of our simulations refer to \emph{career paths} of individual agents.
From these, we have to reconstruct an aggregated network of mobility. % as described in Section~\ref{sec:mobility-network}.
Our simulation results for the network topology are reported for these simulated networks.

% The second two distributions that we consider are computed by taking a scientists-centric perspective.
% We look at the distribution of distances travelled by scientists and the probability that scientists move depending on their academic age.

% To validate our model, we verify that it is able to reproduce four empirical distributions.
% The first two distributions that we consider are computed by taking a network perspective.
% Precisely, by considering the city as nodes and the scientists' movements across cities as links, we can construct a network mobility network.
% We compute the distribution of clustering coefficients and path lengths for both the empirical network and the simulated networks with optimal parameters.
% In Fig.~\ref{fig:net-valid_1}, we report the comparison between the empirical (red) and simulated (blue) distributions:
% in Fig.~\ref{fig:net-valid_1} (a) we compare the distributions of clustering coefficients
% and see Fig.~\ref{fig:net-valid_1} (b) we compare the distributions of path lengths.
% %two distributions: in red the empirical and in blue the simulated.

In conclusion, we report that our agent-based model captures the different features of the empirical data well, both on the scientists' and the network level, without using direct information from these for the calibration.
% In Fig.~\ref{fig:traj-valid}, we report the empirical (red) and simulated (blue) distributions and find a good match between them.
% By checking that we correctly reproduce both system properties and scientists-centric properties, we find that our model captures real-world patterns of scientists mobility.

% Note that during this calibration procedure, we do not use any real distance or any dyadic properties between cities.
% By this, we argue that any correctly reproduced network properties and mobility pattern will depend on our model

% \clearpage

\section*{Discussion}
\label{sec:discuss}

This paper provides several results with relevance for both the empirical and the theoretical understanding of the global mobility of scientists.
As a novel contribution, we introduce the concept of a \emph{geographical career path} of an individual scientist, which can be extracted from data.
Using records of 3.5 Mio scientists, we provide a statistical analysis of such career paths, that later form the basis for comparison with our model, on the scientists' level.
Aggregating over these career paths, we are further able to reconstruct the \emph{world network of scientists' mobility}, with cities as nodes and inflow/outflow of scientists as links.
With this, we reveal the patterns of scientists' mobility on two levels: the level of an individual scientist (age, relocation distance), and the level of cities forming a global network, which is a new empirical insight.

The most important contribution, however, is an agent-based model that allows reproducing these empirical findings on both the scientist and the city level.
In our model, we assume as most relevant factors geographical distances, academic importance, and selectiveness of cities.
The model uses as input for the initial conditions only variables that can be proxied by the available data.
In particular, academic importance, denoted as the fitness of agents, is proxied from the available publications of scientists.
The fitness of locations, another ingredient of the model, is then obtained by averaging over the fitness of agents at the particular location.

The agent-based model succeeds with simple assumptions for the relocation of agents.
Agents rank all locations according to their fitness and their distance to the current location.
However, they do not decide whether to move.
This choice is made by the locations using the information on the fitness of the agents and capacity constraints.
In essence, this poses a matching problem and can be related to similar problems discussed in the literature.

Our agent-based model only considers two free parameters, which need to be calibrated against the available data:
$b$ weights the spatial distance between the current location of an agent and any other location, $s$ weights the selectiveness of locations when accepting agents that have a fitness below the location's fitness.
We find as \emph{optimal parameters} $(s^{\mathrm{opt}}, b^{\mathrm{opt}}) = (0.5, 0.5)$.
These parameters are maximally different from 0 or 1 and indicate that both selectiveness and distances are essential to reproduce the empirical mobility patterns.
$b^{\mathrm{opt}}=0.5$ characterizes the \emph{supply side}, i.e., the ranking of locations by the \emph{agents}.
A location $A$ needs to have $\sqrt{n}$ times the fitness of another location $B$ if it is $n$ times further away, to be equally attractive for an agent.
$s^{\mathrm{opt}}=0.5$ characterizes the \emph{demand side}, i.e., the ranking of agents by \emph{locations}.
Provided there are sufficient openings available, an agent with a fitness ratio $f_i/F_k=0.5$ is accepted more than 2 out of 3 times,
and an agent with $f_i/F_k=0.25$ still has a 50\% chance to be accepted.

\begin{figure}
    \centering
    \includegraphics[width=0.8\textwidth]{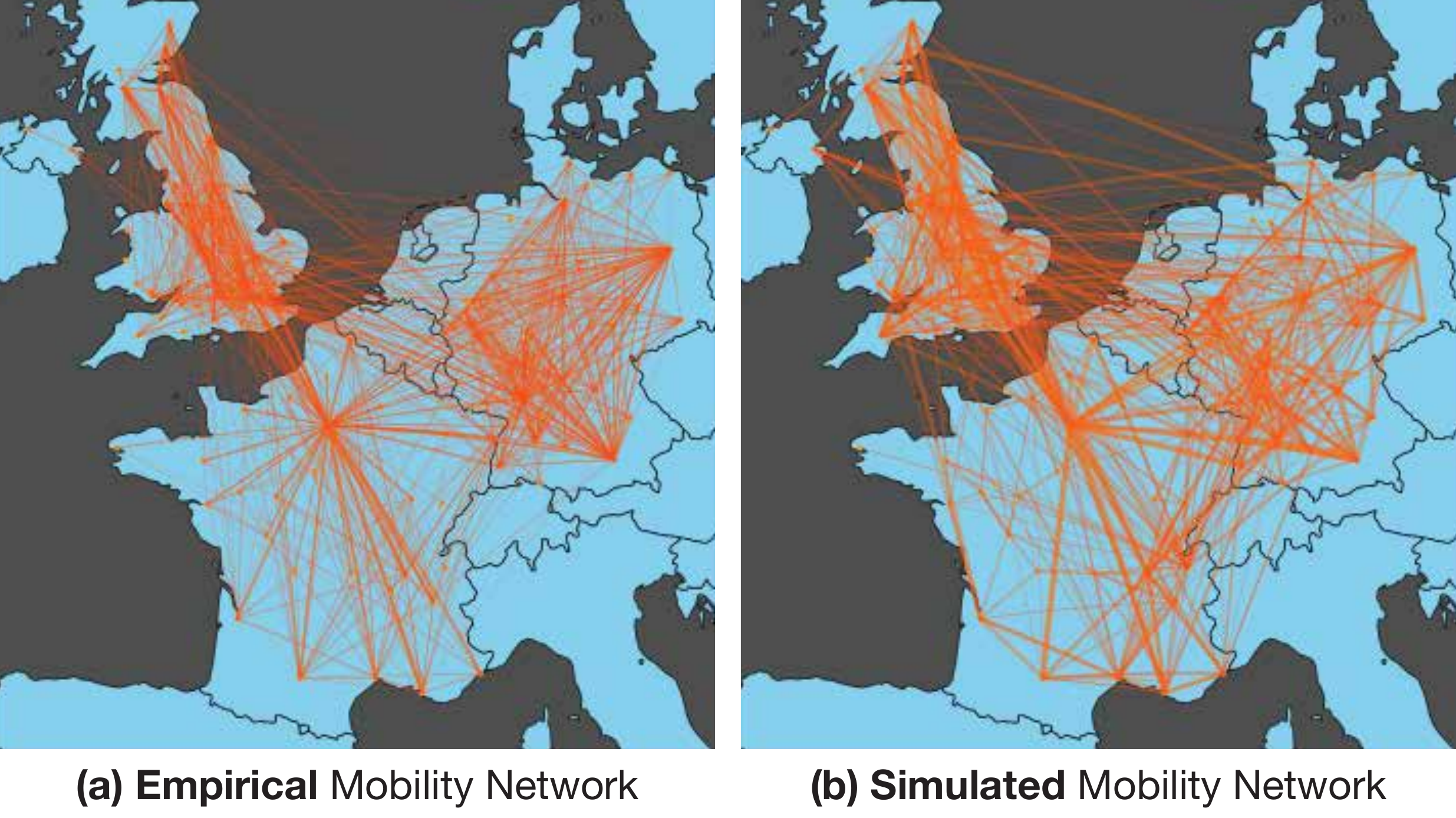}
    \caption{\emph{Empirical and Simulated mobility networks for France, Germany and UK.}
    The empirical network (a) depicts the flows between cities, the thickness of links indicates their magnitude. The map in (b) depicts one realization of the ABM with optimal parameters.
    }\label{fig:geo-valid}

\end{figure}

Using the model calibrated with the optimal parameters, our simulations match the available empirical data quite well.
This is remarkable because the model does \emph{not} include many factors which arguably play a role in the relocation decision.
In other words, only using minimal assumptions about the supply (scientists) and demand (cities), and a simple matching mechanism, we are able to capture emergent features of the scientist mobility network.
Some minor differences between the simulated and the empirical distributions become only noticeable if we plot the network of scientists' mobility on the European scale, as shown in Figure~\ref{fig:geo-valid}.
We observe that the empirical network in Figure~\ref{fig:geo-valid}(a) has more pronounced hubs than the simulated network shown in Figure~\ref{fig:geo-valid}(b).
Specifically, in the empirical network, significantly more French cities are linked to Paris than in the simulation.

% In Sect.~\ref{sec:discuss}, we further discuss this bringing to the attention also the limitation of our model.
% The only notable differences are the first peak that is more pronounced in the simulations, and the empirical distributions have heavier tails.
% To study the empirical mobility patterns of scientists, we have started reconstructing the geographical trajectories of scientists.
% We have represented these trajectories using a network perspective, and we have proposed an agent-based model to reproduce the network.
% After the calibration of the model we find that the simulated network is able to reproduce distance accurately travelled,
% age distribution, local clustering and number of moves (see Fig.~\ref{fig:net-valid_1} (a) and (b) and Fig.~\ref{fig:traj-valid}(a) and (b)).

Finally, we stress that more factors are influencing the relocation choices of scientists than explicitly covered in our model.
For example, quality of life, better networking opportunities or higher salaries might be relevant factors here.
The more remarkable is the fact that our model, even at this level of detail, works considerably well.

In summary, we have provided the first agent-based model reproducing the mobility of scientists.
In a data-driven approach, our model has been calibrated and validated against data, and we have found a remarkably good match between simulations and empirics.
We show that minimal decision rules capture many complex features of the observed mobility of scientists.
Besides, we have quantified the relative importance between geographical distances and academic attractiveness from the perspective of a scientist trying to relocate.
% (one paragraph about outlooks)
% Reproduce finding number 1 of your paper, i.e. ``moves from east to the west are mostly green''.

% Analysing and reproducing temporal correlations in scientist trajectories.
% We do not find any significant correlations when aggregating all the trajectories, but when sub-sampling by field of specialisation we might.
% This could be captured by introducing more heterogeneous agents.

\section*{Materials and Methods}

\label{sec:mm}

\paragraph{Extracting individual career paths of scientists. }
\label{sec:available-data}

For our work, we use the {MEDLINE} database, the largest open-access bibliographic database in the life sciences.
Our analysis is based on two datasets provided by Torvik and co-authors, namely \texttt{MapAffil} \cite{Torvik2015} and \texttt{Author-ity} \cite{Torvik2009}, which have been extracted from {MEDLINE}.
\texttt{MapAffil} lists for each {MEDLINE} paper and each scientist the \emph{disambiguated city names} of the listed affiliation (37,396,671 city-name instances).
It further gives a unique identifier as well as the geo-coordinates of each city.
\texttt{Author-ity} contains the \emph{disambiguated scientist names}, linking them to their respective publications.
Combining these two sources of information about geo-coordinates and time allows us to construct the geographical ``career path'' of scientists, using the approach of~\cite{verginer2020global}.

Formally we denote a career path of a scientist $i \in N$ as a sequence $p_i$, for example, $p_i=\{A_{t_0},$ $A_{t_1},$ $A_{t_2},$ $B_{t_2},$ $B_{t_4},$ $C_{t_5},$ $C_{t_6}\}$.
$A$ denotes the city as defined by its geo-location $R_{A}=(X,Y)$ where $X$ gives the latitude and $Y$ the longitude according to the
data from \texttt{MapAffil}.
The subscript $t_{0}$ refers to the year scientist $i$ was affiliated in the respective city, according to the career path data obtained.
For more information about the data used, see the SM.
% We illustrate a career path in Figure~\ref{fig:move_example}.
% Note that due to the time resolution of one year, a scientist may have multiple publications as well as multiple locations in the same year.
% % In Figure~\ref{fig:move_example}, we illustrate this case as we observe $A$ and $B$ simultaneously at $t_2$.
% In such a case, we choose the location that has been observed more often in the window.
% If two locations are equally frequent, then we select the one that is closest to the move year $t$.
% In the case of perfect ties, one location is chosen at random.

\paragraph{Determining locations in geographical space.  }

Defining the boundaries of a city is a central problem in urban studies.
A standard definition available for US cities is the ``Metropolitan Statistical Area'' (MSAs)~\cite{Census2018}.
However, as the name suggests, this definition is \emph{not} available outside the US.
Therefore, to identify cities, we rely on the definition of ``location'' provided by Google Maps.
This definition reflects administrative boundaries, which are not perfect substitutes.
As argued by \cite{eeckhout2004gibrat, bee2013size} natural and administrative definitions follow different size distributions.
However, because we do not argue about the size distribution of cities, this is not a crucial concern.

\paragraph{Determining the free model parameters.  }

Parameter $b$ weights the impact of spatial distances on the individual rankings of agents.
$b=0$ would imply that distances do not play a role in relocation preferences; with $b=1$, a location $A$ which is twice as far away as location $B$ needs to have twice the fitness of $B$ ($F_A=2*F_B$) to be equally attractive.
In general, a location $A$ which is $n$ times as far away as $B$ must have a fitness $F_A = F_B*n^{b}$ to be as attractive as $B$.

Parameter $s$ weights the flexibility of locations to still accept agents with a fitness less than the fitness of the location.
For $s=0$, the probability of a location to accept agents is independent of their fitness and always equal to 1 ($p(K,i)=1$); with $s=1$ the probabilty to be accepted is equal to the ratio between the agent's and location's fitness ($f_i/F_k$), e.g., an agent with $f_i=1/2 F_k$ will be accepted with probability $1/2$.
In the case of $s>1$, an agent is accepted by a location with a probability smaller than their fitness ratio ($p(K, i) < f_i/F_k$).

\paragraph{Calibration procedure. }

To \textit{calibrate} the model parameters $b$, $s$, we use an established approach in agent-based modeling~\cite{vaccario2018quantifying},    machine learning~\cite{rokach2005top, lee2010improving, bergstra2012random}    and computer simulations in general~\cite{law1991simulation}.
It combines two elements: (a) a grid search and (b) a performance score.

The grid search consists of an exploration of the (low dimensional) parameter space through computer simulations.
For $b$ the values $\{0.005$, $0.01$, $0.05$, $0.1$, $0.5$, $1.0$ ,$5.0\}$ are considered, for $s$   the values $\{0.05$, $0.1$, $0.5$, $1.0$, $5.0$, $1.0$ ,$10.0\}$.
   % Recall that the model parameters are \textit{only} two: $b$ and $s$.
   % The former weights the importance of distances %(see eq.~\ref{eq:gravity})
   % ,
   % while the latter is the selectiveness of cities %(see eq.~\ref{eq:prob})
   % .
For each parameter combination, we obtain from the simulations two distributions for the \textit{inflow} and \textit{outflow}. %, as shown in Figure~\ref{fig:calibr-distr}(a,b).
To determine the optimal combination of $(b,s)$, we compare the simulated and the empirical inflow and outflow distributions.
For this comparison, we use a performance score based on the Kolmogorov-Smirnov(KS) statistic~\cite{kolmogorov1933sulla}.
   % We choose as optimal parameters that allow reproducing two distributions.
   % The first is the distribution of scientists' outflow per city, i.e. the number of scientists moving away from cities.
   % The second is the distribution of scientists' inflow per city, i.e. the number of scientists moving to cities.
   % To decide which are the parameters better reproducing these two distributions, we introduce a performance score.
Precisely, for each combination of parameters $(b, s)$, we compute the KS-statistic between the empirical and simulated distributions of inflow, $D_1(b,s)$, and of outflow $D_2(b,s)$.
We then define the performance score as $1 / ( D_1(b,s) \times D_2(b,s) )$, such that the optimal combination $(b^{\mathrm{opt}}, s^{\mathrm{opt}})$ maximizes this score.

%%% OLD description of the optimal score
% To decide which are the parameters better reproducing these two distribution, we use the following expression:
% \begin{equation} \label{eq:optimal}
%     p^{\mathrm{opt}} =(b^{\mathrm{opt}}, s^{\mathrm{opt}}) = \argmin_{b,s}\frac{1}{N}\sum^N_k{D_1(b,s) \times D_2(b,s)}
% \end{equation}
% where $N$ is the number of simulations,
% $D_1(b,s)$ is the Kolomogorov-Smirnov statistic between the empirical and simulated distributions of city outflow.
% $D_2(b,s)$ is the Kolmogorov-Smirnov statistic between the empirical and simulated distributions of city inflow.
% In other words, the performance score for each combination of parameters is the average product between $D_1(b,s)$ and $D_2(b,s)$.
% The optimal combination of parameters is the one that minimises this score.

% In Fig.~\ref{tab:opt_param_and_heatmap}, we report a heat-map showing the exploration of the parameter space.
% For each combination of parameters, 10 simulations are run.
% We find as \emph{optimal parameters} $p^{\mathrm{opt}} =(s^{\mathrm{opt}}, b^{\mathrm{opt}}) = (0.5, 0.5)$.
% This means that both selectiveness and distances better reproduce the empirical data when they give a sub-linear contribution.

From the calibration procedure, we find \emph{optimal parameters} $(s^{\mathrm{opt}}, b^{\mathrm{opt}}) = (0.5, 0.5)$.
The comparison between the empirical and the simulated distributions is shown in Figure~S6(a,b) in SM.
The close match demonstrates that our model is correctly calibrated. % as the empirical and simulated distribution overlap quite well.
  % Some differences are discussed in Sect.~\ref{sec:discuss}.

\paragraph{Simulation initialisation.  }

At the beginning of the simulation, we populate the cities at only 80\% capacity, which means that we initiated the simulation with 22,000 agents, and ca. 300 locations.
As the starting year $t=0$, we take 2000.
From each city, we take its geographical position and the number of scientists in the year 2000.
We assign these quantities to locations to characterize their $R_{K}$ and $N_{K}(t=0)$.
The initial fitness value of a location, $F_{K}(t=0)$, is determined by averaging over the fitness values of those agents based in the given city in 2000.

%, its academic impact \textbf{WHAT IS THIS?}
%
%$(f,t_a)$ pairs for the agents.
From each scientist, we obtain its geographical position (in a given city), his/her academic impact, and the years of activity as of the year 2000.
We assign these quantities to agents to characterize their $r_{i}(t=0)$, $f_{i}(t=0)$ and $y_{i}(t=0)$.
The academic impact is proxied by the papers that a scientist has authored in the two years prior.
As above described, we assign to each paper a score equal to the impact factor of the journal where it was published divided by the number of co-authors.
Then, for each scientist, we sum the scores of the papers he/she has co-authored between $1998$ and $2000$.
This defines the starting fitness of agents, i.e., $f_i(t=0)$.
We then run the agent-based model using parallel updates of all agents per time step.%, taking as evolving quantities only the values of $N_{K}(t)$ into account.

%The number of scientists in cities is assigned to the number of scientists hosted in cities.
%The masses are computed by averaging impact scores of scientists hosted in cities.
%The positions of locations corresponding to the geographical position of cities.
%The $(f,t_a)$ pairs are academic impact and academic age of real scientists.
% \begin{table}[]
%     \centering
%     \footnotesize
%     \begin{tabular}{|c|c|c|}
%     \hline
%     \textbf{Input data}                               & \textbf{Model feature}                                                & \textbf{Entity}   \\
%     \hline
%     Number of scientists in cities                    & Carrying capacity of locations              & \texttt{location}  \\
%     Academic impact of scientists aggregated over city & Starting masses of locations                & \texttt{location}  \\
%     Geographical positions of cities                  & Positions of locations                      & \texttt{location}  \\
%     Academic impact and academic age of scientists     & Starting fitness and academic age of agents & \texttt{agent}     \\
%     \hline
% \end{tabular}
% \caption{Data used as input together with their respective model features.}
% \label{tab:input}
% \end{table}
%We summarize in Table~\ref{tab:input} the data used as input.

\paragraph{Simulating the entry and exit dynamics of agents. }

Our empirical analysis finds that the number of scientists is almost constant in six years time windows (see Figure~S3 in SM).
Thus, the total number of agents is almost constant during our simulations.
Specifically, the number of new agents $n_n$ is proportional to the number of removed agents $n_r$ at the previous time step.
We sample $n_n$ from a Gaussian distribution with mean $n_r$ and standard deviation $\sigma = n_r (0.1/2)$.
%Also, we do not model the growth or decline in the number of scientists.
%We focus only on their mobility.

\paragraph{Simulating the matching problem.  }
To match agents with locations,
we first create a ranking of the agents according to their fitness.
Starting from the agent with higher fitness, we look at its top five preferred locations.
If one of these locations accept the agent, we move it there.
When an agent $i$ has moved to a new location $K$, we update its position vector, $r_i(t+1) = R_{K}$, and keep its fitness constant, $f_i(t+1) = f_i(t)$.
Then, we consider the second agent in the ranking and keep trying to match it to a new location.
With this approach, we ensure that agents relocate to their preferred locations if they are accepted.
Also, since we first try to match agents with higher fitness, locations obtain agents with higher fitness, i.e., their preferred ones.

\section*{Data Availability}

The raw XML data on all MEDLINE articles are available for download from the NIH at

\begin{itemize}
	\item \url{https://www.nlm.nih.gov/databases/download/pubmed_medline.html}
	\item \url{ftp://ftp.ncbi.nlm.nih.gov/pubmed/baseline}
\end{itemize}

The disambiguation of authors (Authority) and affiliations (MapAffil)~ has been obtained from

\begin{itemize}
	\item http://abel.lis.illinois.edu/downloads.html
\end{itemize}

Access to this resource can be requested for free from the maintainers through the online form on the same page.
Note that due to an agreement with the providers of Authority and MapAffil, these datasets may only be shared by requesting access through the previously mentioned online form.

We make the aggregated mobility network at city level available with no individual identifying information through figshare after publication.

%\section*{Acknowledgements}

\printbibliography

\section*{Additional Information}

\subsection*{Competing Interests}
The authors declare no competing interests.

\subsection*{Author Contributions}

G.V., L.V. and F.S. conceived and designed the study and formulated the mechanisms.
G.V. and L.V. have prepared the figures and illustrations.
G.V. and L.V. have carried out the simulation and statistical analysis.
G.V., L.V. and F.S. have written the manuscript.
All authors have read and approved the final text.

\end{document}

\title{Supplementary Information for:\\ Reproducing scientists' mobility: A data-driven model}
\titlealternative{Supplementary Information}

\author{Giacomo Vaccario$^{1}$, Luca Verginer$^{1,2,\ast}$, Frank Schweitzer$^{1}$}
\authoralternative{G. Vaccario, L. Verginer, F. Schweitzer}
\address{\bigskip \small
$^1$ETH Z\"urich, Chair of Systems Design, Department of Management, Technology and \\
Economics, Weinbergstrasse 56/58, CH-8092 Z\"urich, Switzerland \\
$^2$IMT School for Advanced Studies Lucca, AXES Lab, piazza S. Francesco 19, IT-55100 Lucca, Italy \\
\bigskip
$^\ast$Corresponding author: fschweitzer{@}ethz.ch
}

\maketitle

\clearpage

\section{Supplementary Information}

\subsection{Data}
We use two datasets extracted from MEDLINE, which is the largest open access bibliographic dataset in the life science\footnote{\url{https://www.nlm.nih.gov/bsd/pmresources.html}}.
The first dataset is \texttt{Author-ity} \cite{Torvik2009} that contains \emph{disambiguated scientist names} and links them to their respective publications in MEDLINE up to 2009.
The second dataset is \texttt{MapAffil} \cite{Torvik2015} which lists for each scientist the \emph{disambiguated city names} of their affiliation ($37\,396\,671$ city-name instances) listed in the MEDLINE publications up to 2015.
Note that \texttt{MapAffil} covers MEDLINE up to 2015, and \texttt{Author-ity} covers MEDLINE up to 2009.
This discrepancy means that we can only use the years up to 2009 when combining the datasets.

By combining the two datasets, we extract for each scientist his/her ``career path''.
An example of such a career path is shown in Table~S\ref{tab:example_medline_record}.
The merged dataset contains a total of of $N=3\,740\,187$ career paths of scientists, which were active between 1950 and 2009, traversing $M= 5\,485$ unique cities.
%To calibrate the ABM in the main manuscript, we restrict the sample to Germany, France, and the United Kingdom from 2000 to 2006.
\begin{table}[htb]
\centering
\caption[Example Affiliation Record]{Example of career path of a specific scientist (LM Shul.).
Only a subset of his/her publications is shown. For each record we have the year of publication, the city of the affiliation and the relative PubMed ID identifying the paper.
}
\label{tab:example_medline_record}
    \begin{tabular}{ccc}
    \hline
    Year &           City &     Pubmed ID \\
    \hline
    $\vdots$ &      $\vdots$ &   $\vdots$ \\
    2000 &      Miami, FL, USA &  11054153 \\
    2000 &      Miami, FL, USA &  10928576 \\
    2000 &      Miami, FL, USA &  10714670 \\
    2000 &      Miami, FL, USA &  10634252 \\
    2001 &  Baltimore, MD, USA &  11763581 \\
    2001 &  Baltimore, MD, USA &  11391746 \\
    2002 &  Baltimore, MD, USA &  15177058 \\
    $\vdots$ &  $\vdots$ &  $\vdots$ \\
    \hline
    \end{tabular}
\end{table}

\begin{figure}
    \centering
    \includegraphics[width=0.80\textwidth]{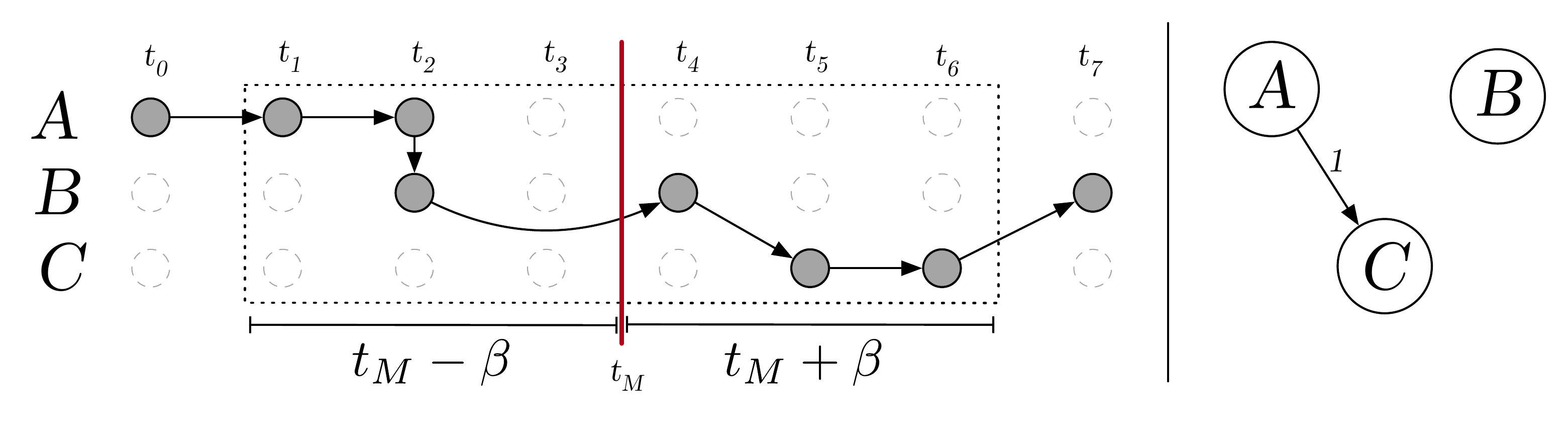}
    \caption{Illustration of procedure to extract movements }
    \label{fig:move_example}
\end{figure}

We have a time resolution of one year.
During a full year, a scientist has often multiple publications, and sometimes is located in multiple cities.
In Figure~S\ref{fig:move_example}, we illustrate this case as we observe locations $A$ and $B$ simultaneously during year $t_2$.
When this happens, we choose the location that has been observed more often in that year.
If two locations are equally frequent, then we select the one that is closest to the move year $t_M$.
In case of perfect ties, one location is chosen at random.
Our procedure follows the approach of~\cite{verginer2018brain}.

\subsection{Entry and exit dynamics}
\label{sec:entry-exit}

\begin{figure}[htbp]
    \centering
    \includegraphics[width=0.70\textwidth]{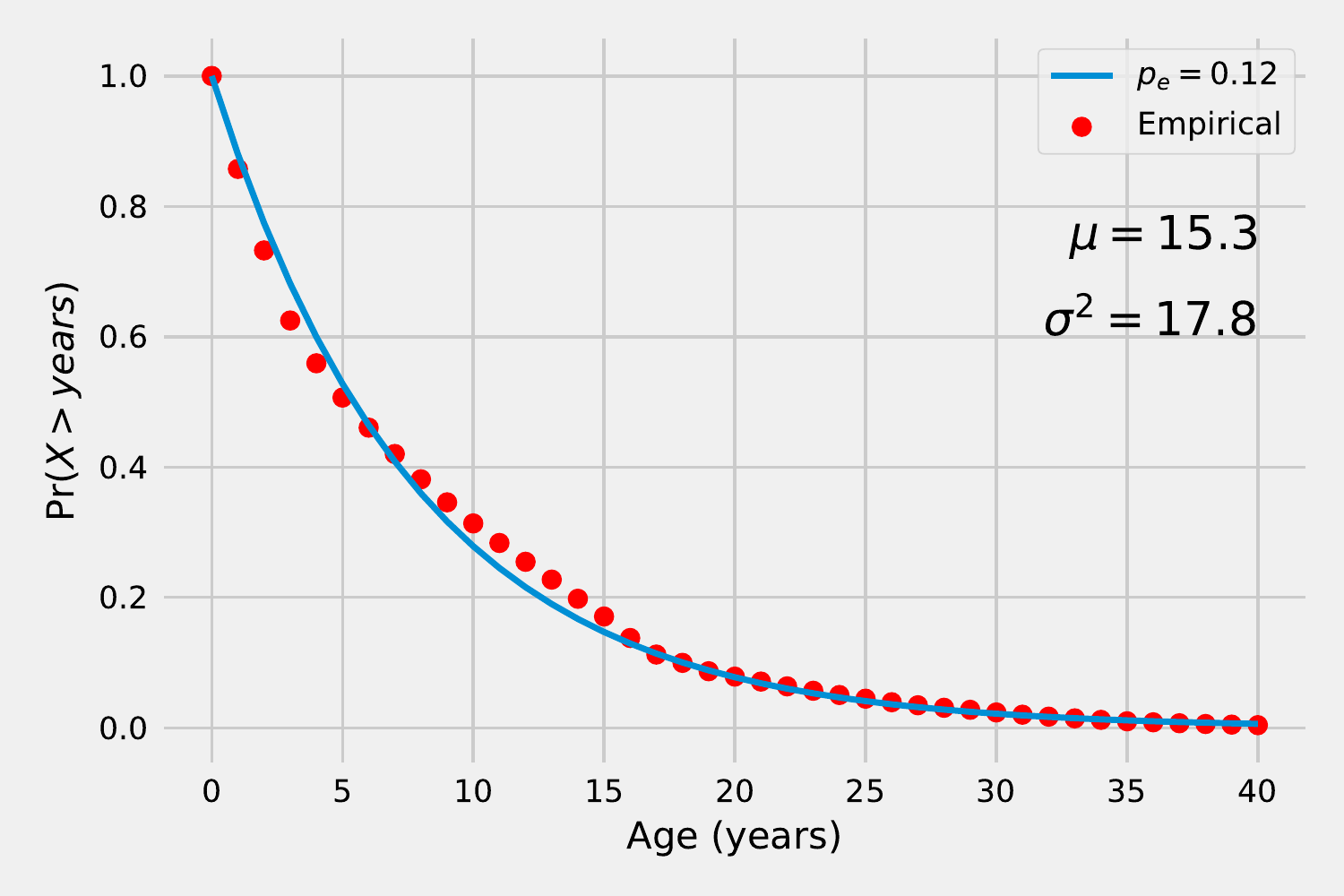}
    \caption{Career Path CCDF.
    The Complementary Cumulative Distribution shows the probability for scientists to remain active, i.e., to publish at least one paper after $k$ years.
    The red markers represent the the empirical CCDF.
    The continuos blue line is the fitted CCDF using an exponential distribution.}
    \label{fig:career-len-distr}
  \end{figure}

\paragraph{Exit probability.} We assume that a scientist has left academia if he/she does not publish for two years in a row.
By counting the number of scientists leaving academia by academic age, we obtain the empirical survival probability (and its complement, the exit probability).
Precisely, in Figure~S\ref{fig:career-len-distr} we show the Complementary Cumulative Distribution Function (CCDF) for the scientists leaving academia by academic age.
Note that this CCDF is well matched by an exponential function with $\lambda \approx 0.1$.
This finding motivate us to chose the exit probability to $p_e=0.1$ in our simulations.

\begin{figure}[htbp]
    \centering
    \includegraphics[width=0.70\textwidth]{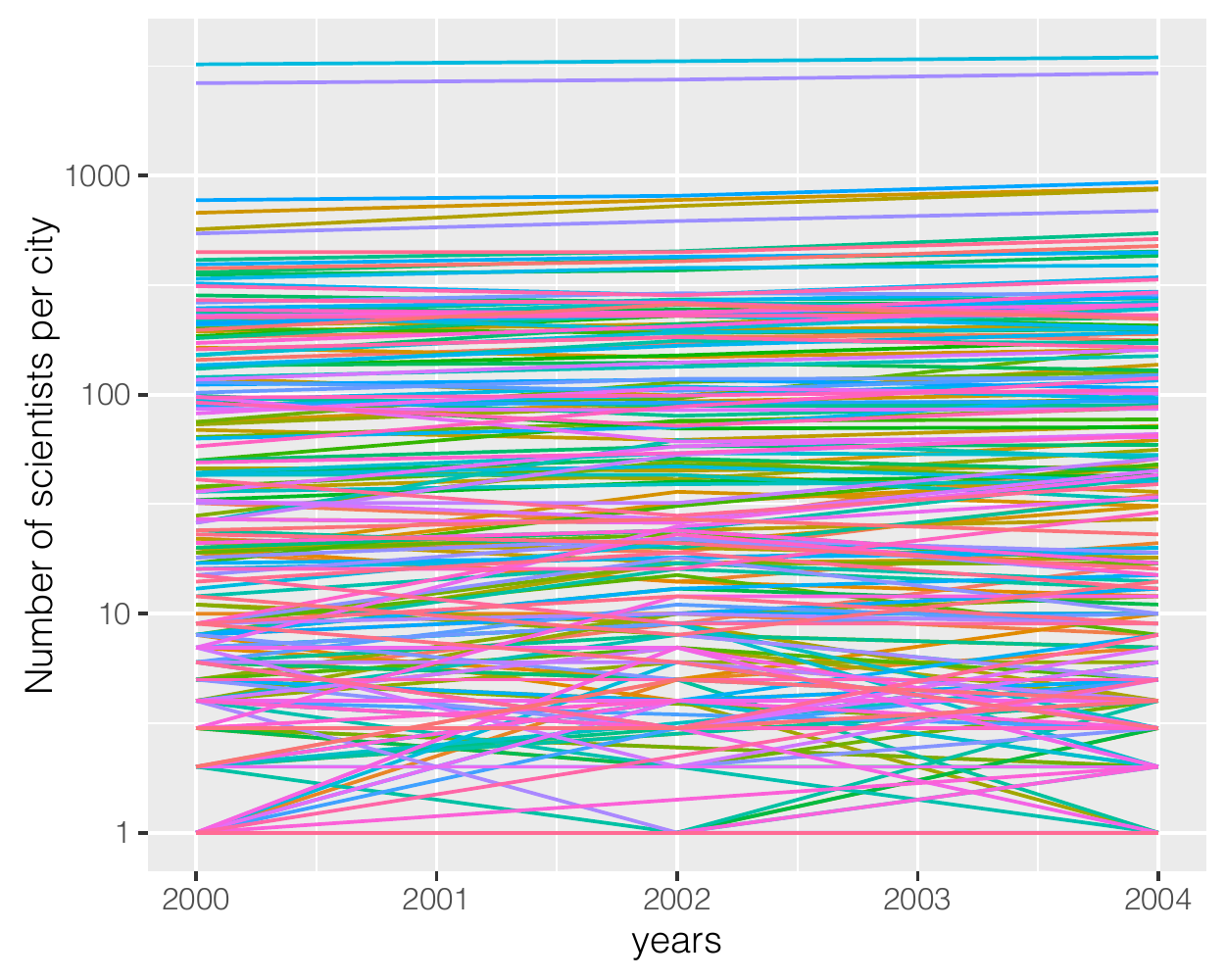}
    \caption{The number of scientists per city in function of time. Each line is the number of scientists per city. We use a log scale in order visualize the heterogeneity of city sizes.
    }
    \label{fig:city-size}
\end{figure}

\paragraph{New agents.}
In Figure~S\ref{fig:city-size}, we report the number of scientists per city in France, Germany, and the UK between 2000 and 2004 (i.e., the countries and period used to calibrate and validate the model).
This number is almost constant, albeit with a slight positive growth trend.
Given the stationarity of city sizes we do not model city growth explicitly.
Tackling a city's scientist population growth is left for future analyses.
% Identifying this trend goes beyond the scope of our analysis, and model presented in our group.
For these reasons, we assume that the number of new agents $n_n$ is close to the number of removed agents $n_r$.
Moreover, we allow for some small fluctuations by sampling $n_n$ from a gaussian distribution with mean $n_r$ and standard deviation $\sigma = n_r (0.1/2)$.

\subsection{Summary of the data-driven modeling}

\begin{figure}
    \centering
    \includegraphics[width=0.99\textwidth]{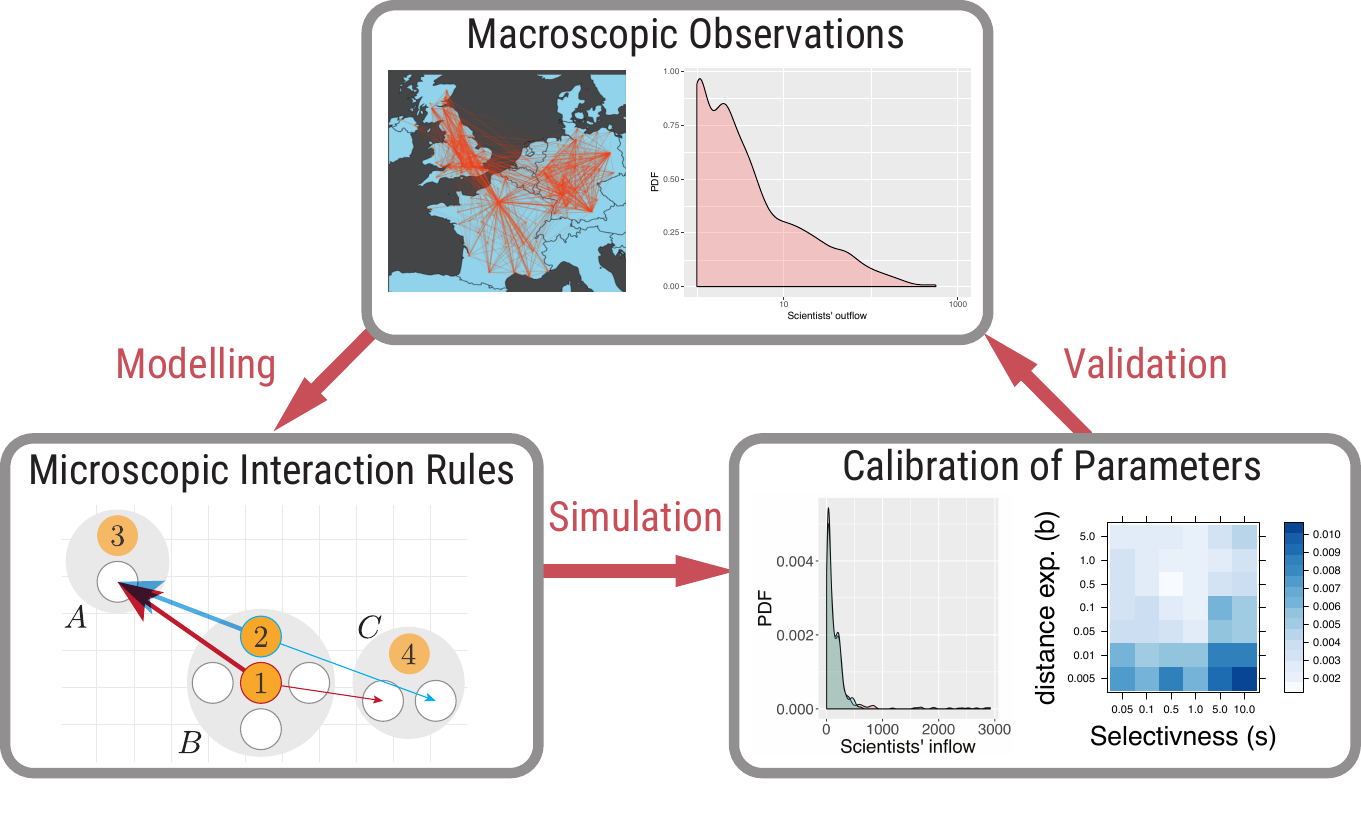}
    \caption{Our modeling procedure is divided in three parts. First, we collect and analyze data about scientists and their career paths across cities. From these data, we extract a set of macroscopic observations both at the scientist and city level (top panel). Second, we use the macroscopic observations to define and inform the model that should reproduce scientists' mobility (bottom left panel). Once we have the model, we simulate and calibrate it using real-data (bottom right panel).
    We call the calibrated model a data-driven model. Finally, we validate the data-driven model by comparing simulation results with the macroscopic observations.}
    \label{fig:graph-abstract}
\end{figure}

In Figure~S\ref{fig:graph-abstract}, we summarize our modeling procedure.
We start from analyzing the data from which we obtain a set of macroscopic observations at the scientist and city level.
These observations range from the academic productivity of scientists to city inflows and outflows of scientists.
Then, we divide these observations in three disjoint sets.
A first set contains observations that act as input data.
Precisely, we use this first set for defining the microscopic rules, the parameters, and the initialization of the model.
We report the observations used as input data in Table~S\ref{tab:input-location} and Table~S\ref{tab:input-agent}.

A second set of observations is used to calibrate the model.
In Table~S\ref{tab:calib}, we report this second set together with the key ingredients of the calibration procedure.
The details of this procedure are in Sect.\ref{app:calib}.
In a nutshell, we obtain the best parameters for the model by comparing simulations with empirical observations.
We call the \emph{calibrated model} a \emph{data-driven model} as both the input variables (e.g., initial conditions, agent and location features) and the parameters are obtained from data.

Finally, the third set of observations is used to validate the calibrated model (see Table~S\ref{tab:valid}).
In the validation procedure, we check whether the model reproduces macroscopic observations both at \emph{scientist} and \emph{network} level.
These observations are the distributions of moved distances, Figure~1~(a), the ``age at move'' distributions, Figure~1~(b), and four distributions of the topological feature shown in Figure~1(c-f).

We note that this is quite an ambitious goal since our model needs to reproduce several \emph{dissimilar} system dimensions (i.e.,\ scientists and intercity) correctly.
If the model can reproduce the described distributions, we have a strong indication that the interaction rules governing scientist and city interactions capture a relevant aspect of the real mobility of scientists.
The information available to the model during calibration does not imply the more complex validation measures.
If we find that the simulated results agree with the empirical validation metrics, it means that the \emph{interaction rules} are the reason for the observed patterns and good validation results.

\begin{table}
    \centering
    \begin{tabular}{|c|c|}
    \hline
    \multicolumn{2}{|c|}{\textbf{Locations}}  \\
    \hline
    \textbf{Input data}                               & \textbf{Model feature}  \\
    Number of scientists in cities                    & Carrying capacity of locations            \\
    Academic impact of scientists aggregated over city & Starting masses of locations              \\
    Geographical positions of cities                  & Positions of locations                      \\
    Almost constant city sizes in 5 years time windows & Constant carrying capacities \\
    \hline
\end{tabular}
\caption{Data used as input together with their respective model features for locations.}
\label{tab:input-location}
\end{table}
\begin{table}
    \centering
\begin{tabular}{|c|c|}
    \hline
    \multicolumn{2}{|c|}{\textbf{Agents}}  \\
    \hline
    \textbf{Input data}                               & \textbf{Model feature}  \\
    Academic impact of scientists     & Starting fitness of agents     \\
    Academic age of scientists     & Starting academic age of agents     \\
    CCDF of scientists leaving academia  & Exit of agents at a rate $\lambda$  \\
    Almost constant city sizes in 5 years time windows & New agents $\approx$ removed agents  \\
    \hline
\end{tabular}
\caption{Data used as input together with their respective model features for agents.}
\label{tab:input-agent}
\end{table}

\begin{table}
    \centering
    \begin{tabular}{|rl|}
    \hline
    \multicolumn{2}{|c|}{\textbf{Distributions used for calibration}}      \\
    \hline
    Network-level            & \begin{tabular}[c]{@{}l@{}}PDF of scientists' inflow (Figure~S\ref{fig:in-out-flow} (a))\\
                                                    PDF of scientists' outflow  (Figure~S\ref{fig:in-out-flow} (b))
        \end{tabular}                               \\
    \hline
    \multicolumn{2}{|c|}{\textbf{Grid search of the parameter space}}      \\
    \hline
    $b$            &  $\{0.005$, $0.01$, $0.05$, $0.1$, $0.5$, $1.0$, $5.0\}$                              \\
    $s$  &  $\{0.05$, $0.1$, $0.5$, $1.0$, $5.0$, $10.0\}$  \\
    \hline
    \multicolumn{2}{|c|}{\textbf{Performance score}}      \\
    \hline
    $\frac{1}{N}\sum^N_k{D_1(b,s) \times D_2(b,s)}$             & \begin{tabular}[c]{@{}l@{}} $D_1$, KS-statistic between the distributions of inflow\\
                                                    $D_2$, KS-statistic between the distributions of outflow\\
                                                    $N$, number of simulations per combination of parameters
        \end{tabular}                               \\
    \hline
    \end{tabular}
    \caption{Summary of the key ingredients of the calibration procedure. We report the distribution used to calibrate the model, the parameter space explored during the grid search, and the performance score used to evaluate the goodness of parameter combinations.}
    \label{tab:calib}
\end{table}
\begin{table}
    \centering
    \begin{tabular}{|rl|}
        \hline
    \multicolumn{2}{|c|}{\textbf{Reproduced distributions}}      \\
    \hline
    Network            & \begin{tabular}[c]{@{}l@{}}PDF of clustering coefficients (Fig.~4~(a))\\
                                                    PDF of shortest path lengths (Fig.~4~(b)) \\
                                                    PDF of degree (Fig.~4~(c)) \\
                                                    PDF of av. neighbour degree (Fig.~4~(d)) \\
        \end{tabular}                               \\
    \hline
    Scientists-centric & \begin{tabular}[c]{@{}l@{}}PDF to observe a scientists to move given its age (Fig.~3~(a)\\
                                                    PDf of distances travelled by scientists (Fig.~3~(b))\end{tabular} \\
    \hline
    \end{tabular}
    \caption{Summary of the distributions used for validation. Note that there macroscopic observations both at the scientist and city/network level.}
    \label{tab:valid}
\end{table}

\subsection{The calibration procedure}
\label{app:calib}
\begin{figure}
    \center
    \includegraphics[width=0.43\textwidth]{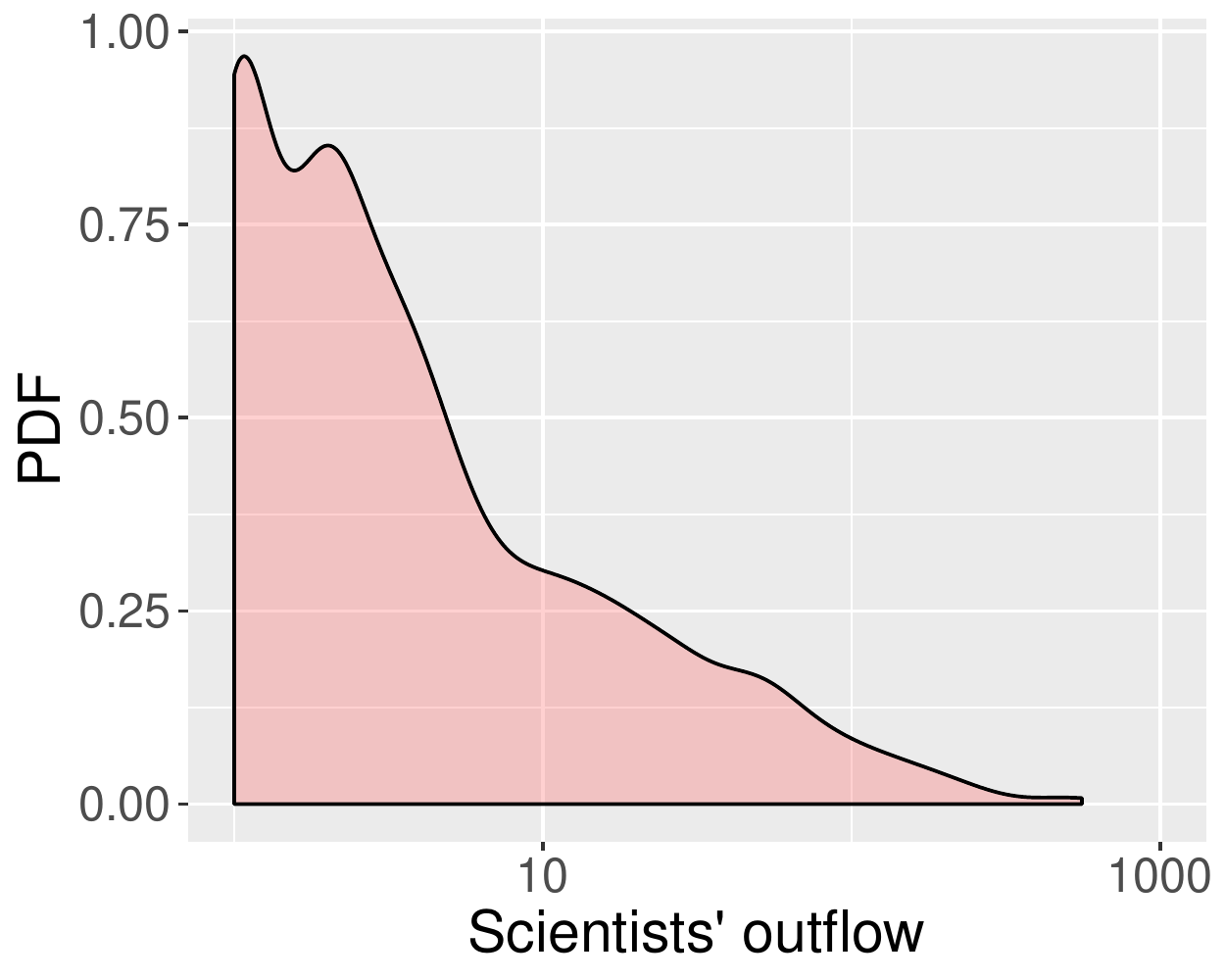}
    \hfill
    \includegraphics[width=0.43\textwidth]{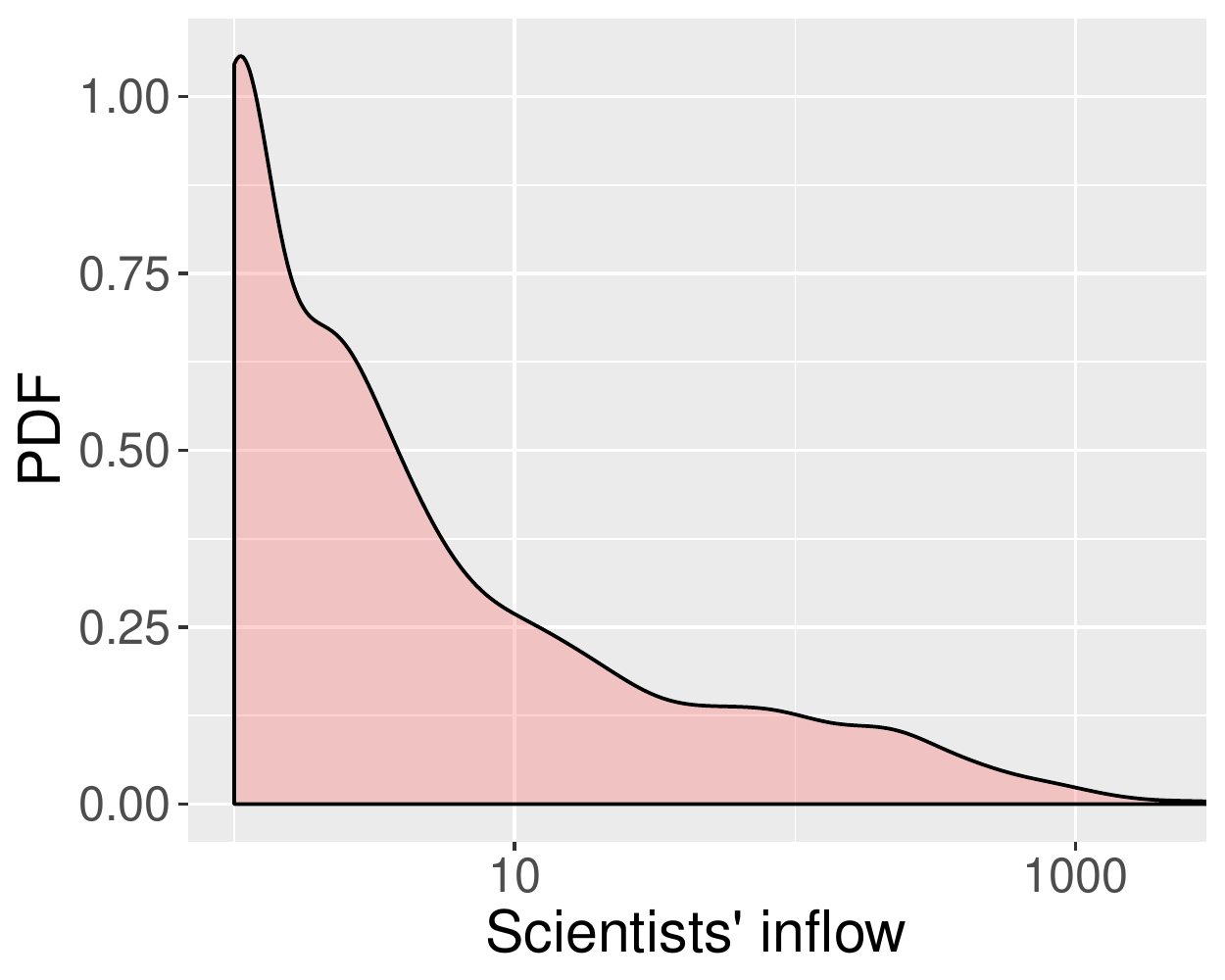}\\
    (a) \hspace{7cm} (b)
    \caption{Distributions of (a) inflow of scientists into any city, (b) outflow of scientists out of any city.
    The $x$-axis is in log-scale.
    }\label{fig:in-out-flow}
\end{figure}
To calibrate the model, we use a set of empirical observation that will not be used for the validation.
In particular, we choose the distribution of in- and outflow of scientists (see Figure~S\ref{fig:in-out-flow}).
As written in the main manuscript, we define for each year $t$ the number of scientists $\Delta N_{K\gets L}(t)$ moving into city $K$ from another city $L$, i.e., the inflow, and the number of scientists $\Delta N_{L\gets K}(t)$ moving out of city $K$ to another city $L$, i.e. the outflow.

Figure~S\ref{fig:in-out-flow}~(a) and (b) show the respective distributions for the aggregated inflow $\Delta N_{K}^{\mathrm{in}}(t)=\sum_{L}\Delta N_{K\gets L}(t)$ of scientists into city $K$ and the aggregated outflow $\Delta N_{K}^{\mathrm{out}}(t)=\sum_{L}\Delta N_{L\gets K}$ of scientists out of city $K$.
The aggregate in- and the out-flow are computed at three different time windows centred at 2000, 2002 and 2004.
This means that each city is considered three times (once in each time window).
We note the left-skew distribution for both quantities, which indicates the \emph{heterogeneous} contribution of cities to the global movement of scientists.

\begin{figure}
    \center
    \includegraphics[width=0.43\textwidth]{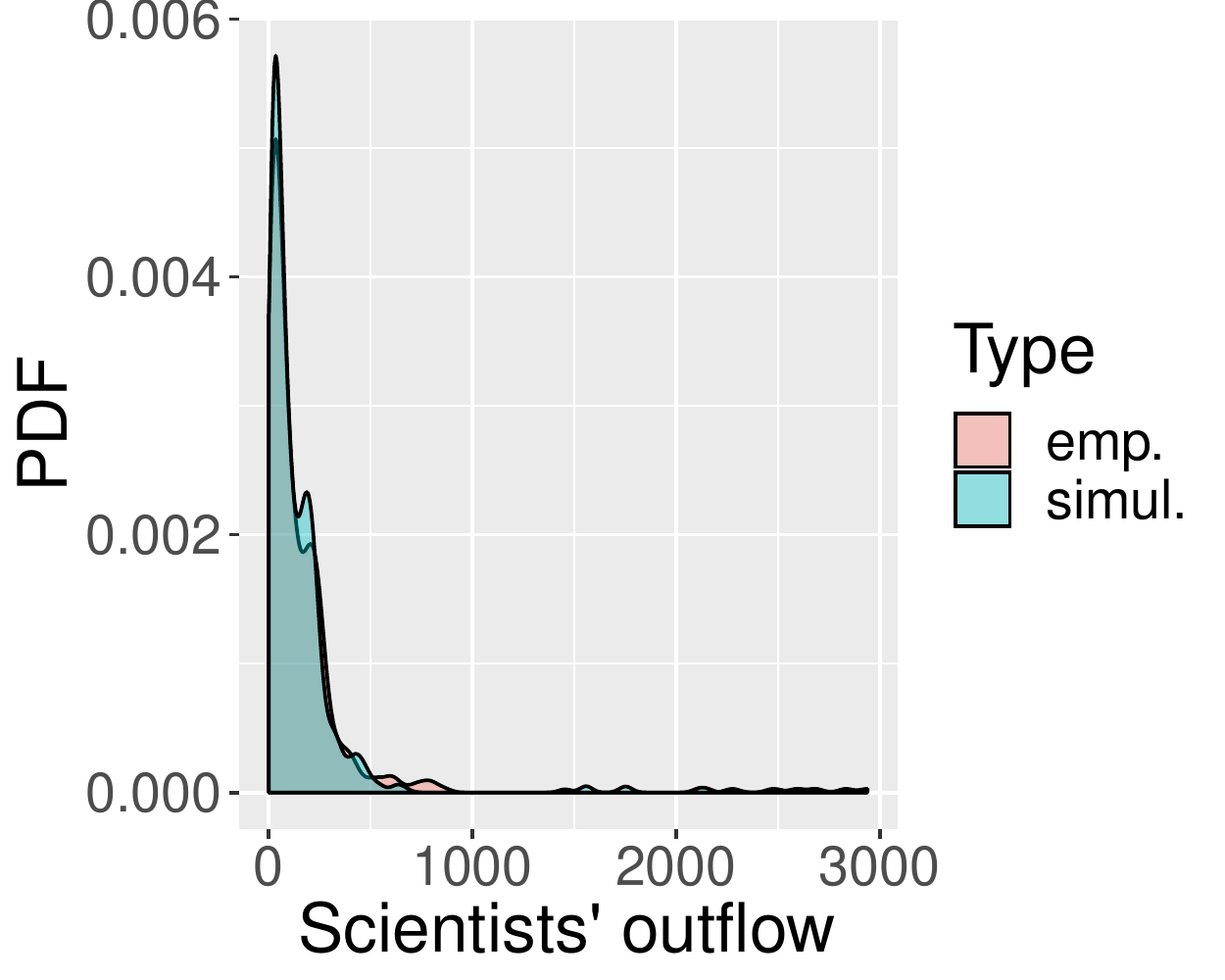}
    \hfill
    \includegraphics[width=0.43\textwidth]{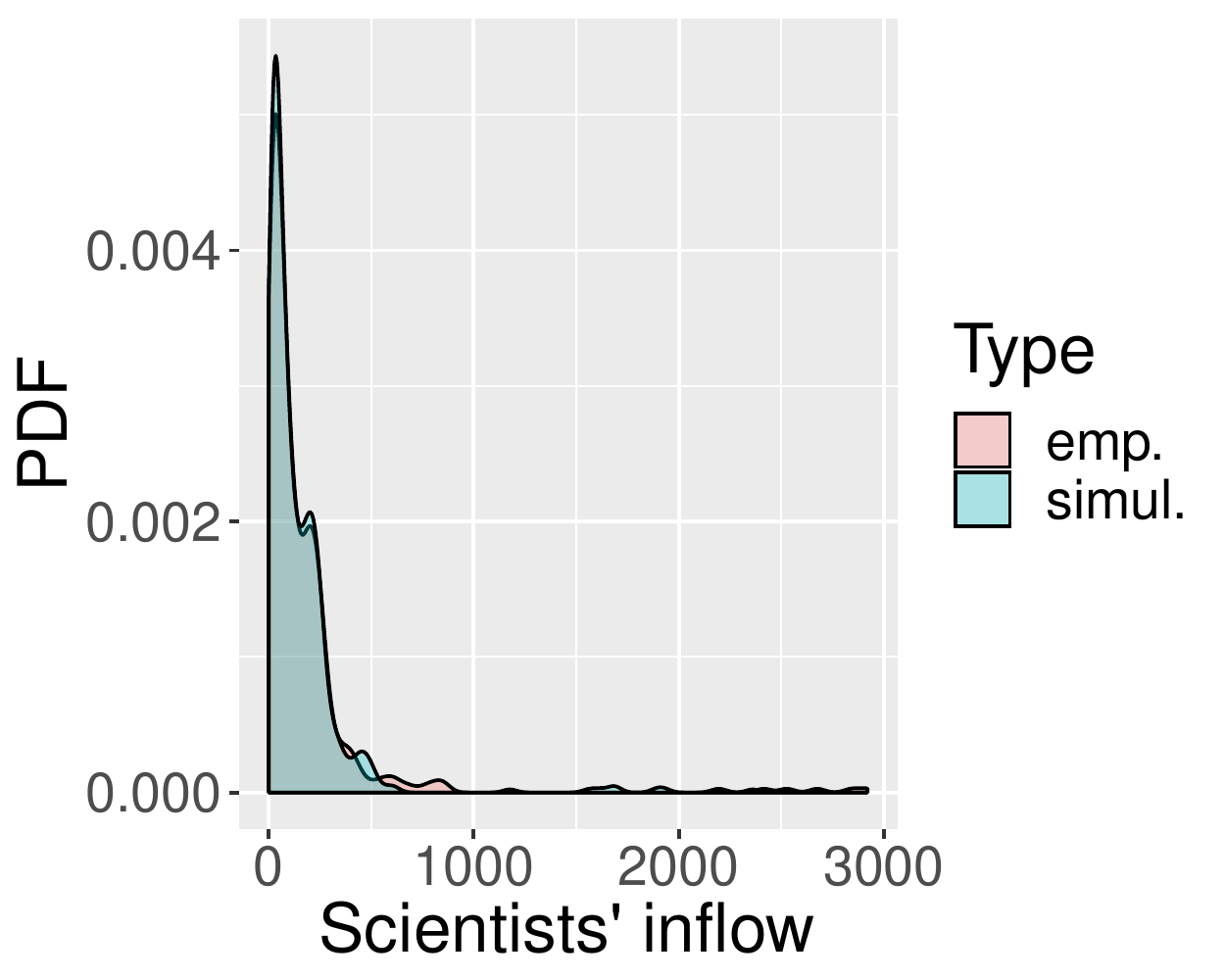}\\
    (a) \hspace{8.7cm} (b)
    \caption{Distributions of (a) inflow of scientists into any city, (b) outflow of scientists out of any city.
      (red) indicates the empirical distributions, (blue) the (optimally) simulated distributions obtained from the calibration of our agent-based model.
    }\label{fig:calibr-distr}
\end{figure}

To calibrate the model, we use the two empirical distributions: the inflow and the outflow distributions shown in Figure~S\ref{fig:calibr-distr}(a,b).
Note that, we calibrate our model considering only cities and scientists present in three countries: France, Germany and United Kingdom.
For this reason, Figure~S\ref{fig:in-out-flow} and Figure~S\ref{fig:calibr-distr} differ.

\begin{figure}
    \centering
    \includegraphics[width = 0.70\textwidth]{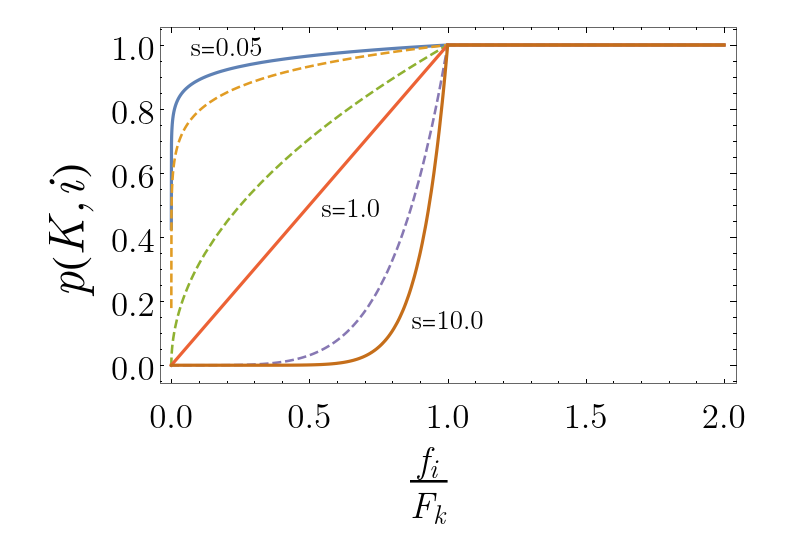}
    \caption{Probability of a location $K$ to accept an agent $i$ in function of the ratio between the agent and location fitness.
    We set $s=0.05$, 0.5, 1, 5, and 10 respectively in blue, yellow, green, red, purple, and orange.}
    \label{fig:decision-prob}
\end{figure}

In the calibration procedure, we start by performing a grid search, i.e., we explore the parameter space defined by our two parameters $b$ and $s$.
For $b$ the values $\{0.005$, $0.01$, $0.05$, $0.1$, $0.5$, $1.0$, $5.0\}$ are considered, for $s$
the values $\{0.05$, $0.1$, $0.5$, $1.0$, $5.0$, $10.0\}$.
Note that with these values of $s$, we explore very different shapes of the acceptance probability $p(K,i)$ defined in the main text.
In Figure~S\ref{fig:decision-prob}, we plot the acceptance probability $p(K,i)$ for five different values of the parameter $s$.
Note that for $s=10$, locations becomes more selective as almost only agent with $f_i\geq F_{K}$ are accepted.
For smaller values of $s$, locations becomes less selective.
For example, for $s=0.05$, locations accept agents almost independently of their fitness.
In Figure~S\ref{fig:decision-prob}, we plot the acceptance probability $p(K,i)$ for five different values of the parameter $s$ and is given by

   \begin{equation}\nonumber
       p(K,i) = \begin{cases}1 &  \frac{f_i}{F_{K}} \geq 1\\
       \left(\frac{f_i}{F_{K}}\right)^{s} & \text{otherwise} \end{cases}
   \end{equation}

Note that at large values of $s=10$, locations becomes more selective and accept almost exclusively agents with $f_i\geq F_{K}$, i.e. only agents which will increase a city's fitness $F_{K}$
For smaller values of $s$, locations becomes less selective.
For example, at $s=0.05$, locations accept agents almost independently of their fitness.

To decide which parameter combination better reproduces the in- and outflow distributions, we use the following performance score:
\begin{equation} \label{eq:optimal}
    p^{opt} =(b^{opt}, s^{opt}) = \argmin_{b,s}\frac{1}{N}\sum^N_k{D_1(b,s) \times D_2(b,s)}
\end{equation}
where $N$ is the number of simulations,
$D_1(b,s)$ is the Kolomogorov-Smirnov statistic between the empirical and simulated distributions of city outflow.
$D_2(b,s)$ is the Kolomogorov-Smirnov statistic between the empirical and simulated distributions of city inflow.
In oder words, the performance score for each combination of parameter is the average product between $D_1(b,s)$ and $D_2(b,s)$.
The optimal combination of parameter is the one that minimizes this score.

\begin{figure}
    \centering
    \includegraphics[clip, trim = 5 65 5 70, width=.4\textwidth]{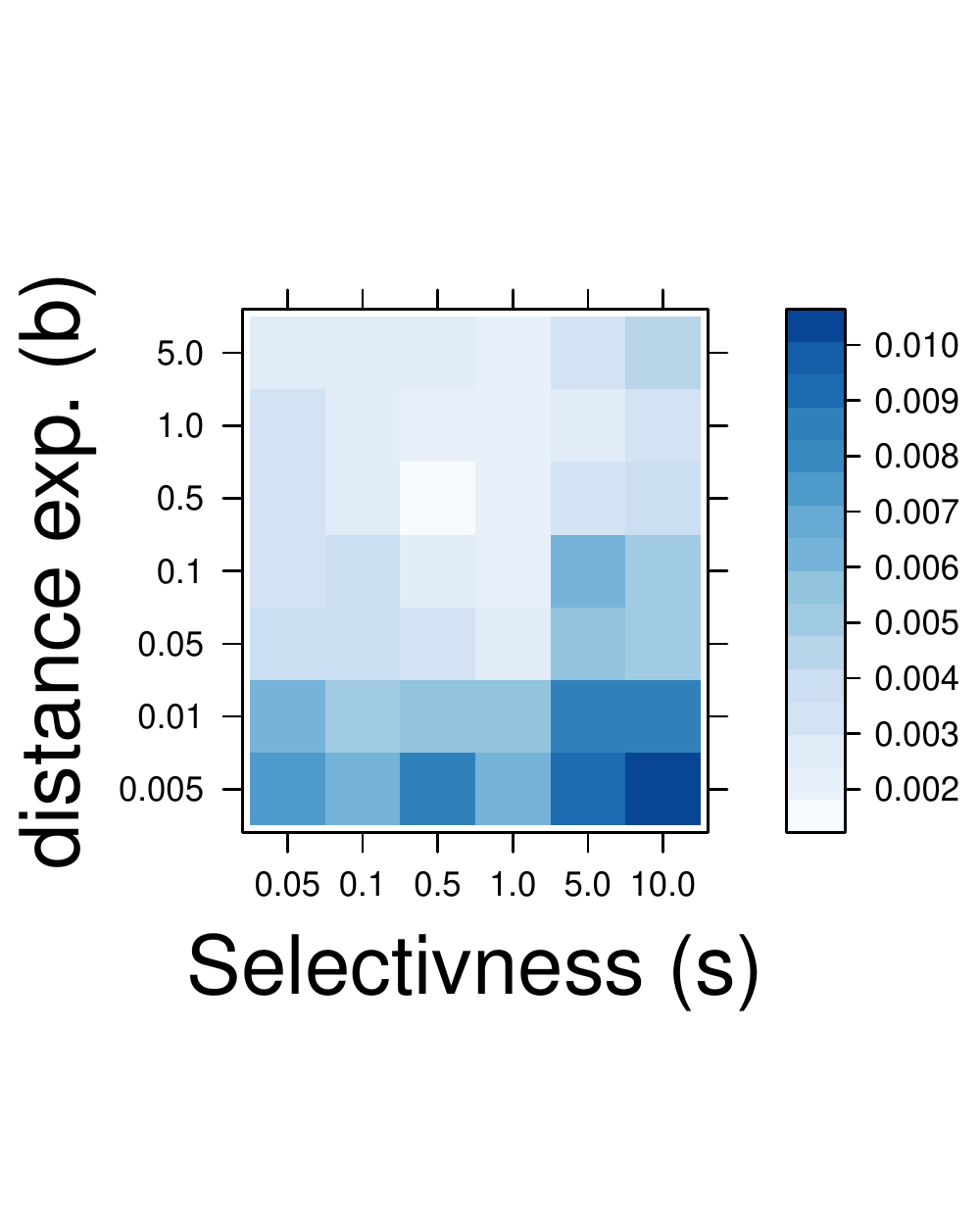}
\caption{The heat-map shows the results of the grid-search on the two parameters $s$ and $b$.
The color of each cell corresponds to a $p$ for a given $(b, s)$ pair as described in eq.~\ref{eq:optimal}.
The optimal parameter pair $(b^{opt}, s^{opt})$ is $(0.5, 0.5)$.
}
\label{tab:opt_param_and_heatmap}
\end{figure}
In Figure~S\ref{tab:opt_param_and_heatmap}, we report a heat-map showing the exploration of the parameter space.
For each combination of parameters, we have performed $N=10$ simulations.
We find as \emph{optimal parameters} $p^{opt} =(s^{opt}, b^{opt}) = (0.5, 0.5)$.
This means that both selectiveness and distances better reproduce the empirical data when they give a sub-linear contribution.

\bibliographystyle{../sg-bibstyle}
\setlength{\itemsep}{4pt}
\bibliography{../biblio_abm}